\documentclass[twocolumn, tighten, astrosymb, twocolappendix]{aastex7}

\usepackage{amsmath}
\usepackage{amssymb}

\usepackage{savesym}
\savesymbol{tablenum}
\usepackage{siunitx}
\restoresymbol{SIX}{tablenum}

\newcommand{\de}{{\rm d}}

\graphicspath{{./}{figures/}}

\begin{document}

\title{The Effects of Magnetic Accretion on the Spatial Extent of White Dwarf Pollution}

\shorttitle{White Dwarf Magnetic Accretion}
\shortauthors{Pham, Taylor, \& Cunningham}

\author[0000-0002-0924-8403]{Dang Pham}
\altaffiliation{AGT and DP contributed equally and are joint first authors.}
\affiliation{JILA and Dept. of Astrophysical and Planetary Sciences, CU Boulder, Boulder, CO 80309, USA}
\email[show]{dang.pham@colorado.edu}

\author[0000-0002-0140-4475]{Aster G. Taylor}
\altaffiliation{Fannie and John Hertz Foundation Fellow}
\affiliation{Dept. of Astronomy, University of Michigan, Ann Arbor, MI 48109, USA}
\email[show]{agtaylor@umich.edu}

\author[0000-0001-7296-3533]{Tim Cunningham}
\affiliation{Center for Astrophysics | Harvard \& Smithsonian, 60 Garden St., Cambridge, MA 02138, USA}
\email{timothy.cunningham@cfa.harvard.edu}



\begin{abstract}
Many white dwarfs are polluted by metals, which are generally understood to be the accreted remnants of a planetary system. 
Modeling these systems typically assumes that the metal concentration is homogeneous throughout the white dwarf's atmosphere. 
However, the magnetic fields of a white dwarf may affect the accretion geometry of the white dwarf via magnetospheric accretion. 
Convection in the white dwarf's photosphere will then transport the metals across the surface, with a structure set by the relative sinking versus spreading timescales. 
In this work, we construct models for the accretion geometry, subsequent spreading, and observed pollution of magnetic white dwarfs. 
We show that the magnetic fields will initially concentrate the pollution into a narrow region of the white dwarf's surface. 
The relative spreading and sinking timescales determine whether the metals become uniformly distributed or remain confined to localized patches.
If the magnetic field and spin poles are misaligned, then patchy white dwarfs exhibit periodically variable pollution signatures, which enable constraints on the patch area.
We explore this model as a possible explanation for the recent detection of periodically variable pollution signatures in magnetic white dwarfs.
Finally, we also demonstrate that the concentration of material due to the magnetic field may lead to systematic underestimates of the mass accretion rate onto these objects.
\end{abstract}


\section{Introduction}

More than \num{1700} white dwarfs have been observed with photospheres polluted by metals \citep[e.g.,][]{Zuckerman2003, Williams2024}, with the consensus that these heavier elements originated from the surrounding planetary system \citep[e.g.,][]{Jura2003, Zuckerman2010}.
This phenomenon is believed to be wide-spread, as up to \qty{50}{\percent} of white dwarfs are expected to be polluted \citep{Zuckerman2003, Zuckerman2010, Koester2014, Manser2024, OuldRouis2024}.
There are many proposed dynamical mechanisms to deliver reservoirs of exoplanetary bodies to the tidal disruption limit, ranging from Oort Clouds under the influence of the Galactic tide to secular chaos in inner planetary systems \citep[e.g.,][]{Alcock1986, Debes2002, Mustill2014,Smallwood2018,Trierweiler2022, OConnor2022,OConnor2023,Akiba2024,Pham2024,Veras2024}.
These pollution reservoirs are typically constrained by the composition of metals observed on white dwarfs, while the dynamical mechanisms must be capable of producing the inferred accretion rates.
Most polluted white dwarf spectra resemble the composition of rocky solar system bodies while inferred accretion rates can span many orders of magnitude \citep{Bauer2019, Putirka2021, Xu2021, Blouin2022}.

Simultaneously, some white dwarfs exhibit strong magnetic fields of unknown origin, with three main candidates: fossil fields \citep{Tout2004, Cantiello2016}, crystallization \citep{Isern2017, Camisassa2024, Fuentes2024}, and binary interactions \citep{Tout2008, Briggs2018, Ferrario2020}.
In a volume-limited spectropolarimetric survey, \citet{Bagnulo2021} found that more than 20\% of white dwarfs within \qty{20}{pc} of the Sun are magnetic, with observed field strengths of \qty{40}{kG} - \qty{100}{MG}. White dwarfs have been discovered to have field strengths reaching hundreds of MG \cite, with the largest known field reaching $B\simeq\qty{900}{MG}$ \citep{caiazzo2021}.
By identifying magnetic white dwarfs with Zeeman splitting in the SDSS \qty{100}{pc} samples, \citet{Moss2025} found two statistically significant subpopulations of magnetic white dwarfs: a young, high-mass, high-field-strength population and an old, average-mass, weak-field-strength population.
Note that the samples in \citet{Moss2025} are limited to field strengths between \qtyrange{1}{100}{MG} due to observational bias.

Since both magnetism and debris disks of tidally disrupted bodies are common around white dwarfs, many previous studies have investigated the process of magnetic accretion onto white dwarfs (e.g., \citealt{Rafikov2011, Metzger2012}).
Magnetic fields as weak as \qtyrange{0.1}{1}{kG} may affect the accreted material \citep{Metzger2012} and may shield the debris disk from highly volatile material \citep{Zhou2024}, which may explain the dearth of observed volatiles in polluted white dwarfs.

Recent observational evidence has shown metal lines that vary in equivalent width with magnetic field strength on the rotational period \citep{Bagnulo2024a}. One possible explanation for this is magnetic accretion on a polluted white dwarf, where metals may be concentrated towards regions of high magnetic field strength \citep{Bagnulo2024a, Bagnulo2024b}, although further work is required to confirm this hypothesis.
Nonetheless, the impacts of magnetic fields on accretion, surface spreading, and the corresponding observational signatures remain undetermined. Of particular interest is the photometric variability and the inferred mass accretion rate of polluting elements.

This work studies the effects of magnetic fields on white dwarf accretion and their observational implications by applying accretion physics of similar systems along with white dwarf simulation techniques \citep{Rafikov2011, Bauer2019, Cunningham2021}.
Specifically, we model the impact of magnetic fields on the accretion of metals, the metal spreading process in a white dwarf atmosphere, and the resulting observational signatures.

We show that the magnetic field strength of white dwarfs will result in significant qualitative differences in the accretion flow, where the magnetic field will truncate the accretion disk once it becomes gaseous.
A simple qualitative model of the white dwarf accretion disk is formulated on the basis of this result.
The magnetic fields will also act to concentrate the pollution into a small region of the white dwarf's surface. 
These polluting metals will only punch through a small fraction of the photosphere before their motion is dominated by collisions with ions.
Once in the photosphere, the existence of a convection zone and the relative sinking versus spreading timescales will determine the structure of the pollution region and the fraction of the white dwarf's surface that is polluted.
The projection of this pollution region from a given viewing angle will have significant impacts on the measured mass accretion rate and the variability of detected metal pollution. 

This paper is structured as follows:  Sec. \ref{sec:accdisk} considers the effects of magnetic fields on the white dwarf's accretion disk and the resulting model for the disk structure, while the effects on the area of the accretion beam are explored in Sec. \ref{sec:spatialdist}. 
We then study the vertical sinking and horizontal diffusion in the magnetic white dwarf's atmosphere in Sec. \ref{sec:diff} and show that the white dwarf's patchiness will remain significant if vertical metal settling is altered or horizontal diffusion is suppressed.
The impact of these results on observed polluted white dwarfs is calculated in Sec. \ref{sec:obs}. 
In Sec. \ref{sec:disc}, we discuss the impact of our results on inferred accretion rates and the caveats of this model, before concluding in Sec. \ref{sec:conc}.

\begin{figure*}
    \centering
    \includegraphics{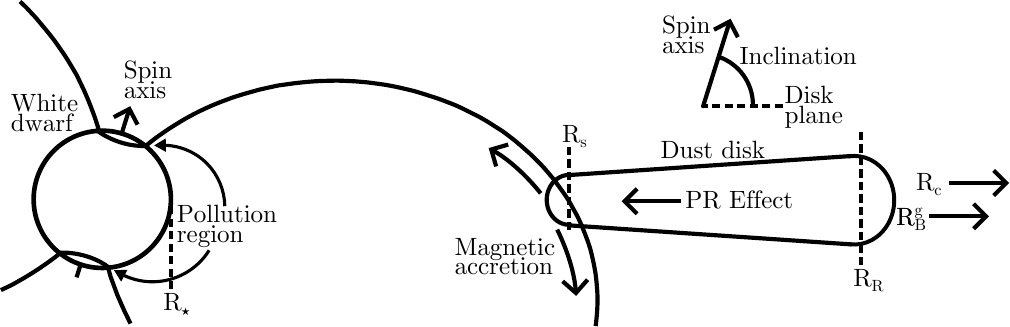}
    \caption{The structure of the white dwarf accretion disk. The magnetic truncation radius $R_B^g$ and the corotation radius $R_c$ are to the right. The disk is not shown to scale. The disk is processed by the Poynting-Robertson (PR) effect from the Roche limit $R_R$. Once the dust reaches the sublimation temperature, at the location $R_s$, it is coupled to the magnetic field lines and begins to accrete onto a white dwarf with radius $R_\star$. The pollution regions will be two symmetric thin bands on the surface of the white dwarf.}
    \label{fig:systemdiagram}
\end{figure*}

\section{Accretion Disk Structure}\label{sec:accdisk}

The material that is accreted onto a polluted white dwarf is expected to be the debris from many sources, ranging from close-in planets to distant exo-Oort Cloud objects (e.g., \citealt{OConnor2022, OConnor2023, Pham2024, Akiba2024, Veras2024}). Once these objects are disturbed onto orbits with perihelia less than the Roche limit, they will be destroyed and form an accretion disk \citep{Jura2003}. This disk will most likely be eccentric and inclined relative to the white dwarf's spin axis \citep{Goksu2024}. This section discusses the structure of the accretion disk and the various forces at work. 

The structure of the accretion disk can be described as follows --- the Roche radius $R_R$ is the outer edge of the accretion disk, the sublimation radius $R_s$ sets the boundary between the gas disk and the dust disk, the gas magnetic accretion radius $R_B^g$ and the dust magnetic accretion radius $R_B^d$ are where the gas and the dust couple to the magnetic fields (respectively), and the corotation radius $R_c$ sets the distance at which material is accreted by magnetic fields, rather than accelerated out of the system. A diagram of the structure of the white dwarf and accretion disk system is shown in Fig. \ref{fig:systemdiagram}. 

The polluting debris disk will be modeled as having two components --- an interior gas disk and an exterior dust disk (see, e.g., \citealt{Rafikov2011, Rogers2025}). Tidally disrupted bodies will feed the dust disk at its exterior edge, which is approximately the Roche radius. This radius is given by (see \citealt{Shu1982})
\begin{equation}
    R_R=R_\star\left(2\frac{\rho_\star}{\rho_b}\right)^{1/3}\,,
\end{equation}
where $\rho_\star$ is the density of the white dwarf, $\rho_b$ is the density of the disrupted body, and $R_\star$ is the radius of the star. At this point, the debris will consist mostly of solid dust particles, with little gas content. Highly volatile gasses will have been removed from the body during its inward trajectory and cleared by magnetic fields \citep{Zhou2024}. Although the dusty material can be approximated as a Shakura-Sunyaev $\alpha$-disk \citep{Shakura1973}, the disk will not be heavily ionized and the dust grains will be relatively large. As a result, the turbulent $\alpha$ viscosity will not be large enough to accrete debris at the rates necessary for observed pollution. Instead, the Poynting-Robertson effect will migrate material inward from the outer (Roche) radius \citep{Rafikov2011, Metzger2012}. In addition, \citet{Okuya2023} showed that the presence of volatiles may enhance the material delivery rate by orders of magnitude above the Poynting-Robertson drag rate.

The boundary between the gas and the dust disk is the dust sublimation radius $R_s$, where the dust is sublimated by the white dwarf's radiation. Following previous work, the dust sublimation radius is (\citealt{Jura2003, Jura2008, Xu2012})
\begin{equation}
   R_s=R_\star\left(\frac{2}{3\pi}\right)^{1/3}\left(\frac{T_\star}{T_s}\right)^{4/3} \,,
\end{equation}
where $T_\star$ is the temperature of the white dwarf, $T_s\simeq\qty{1500}{K}$ is the dust sublimation temperature, and $R_\star$ is the stellar radius. This equation takes the scaled form 
\begin{equation}
    \frac{R_s}{R_\sun}\simeq7.5\times10^{-2}\left(\frac{R_\star}{\qty{0.01}{R_\sun}}\right)\left(\frac{T_\star}{\qty{e4}{K}}\right)^{4/3}\,.
\end{equation}
Interior to this point, the disk is almost entirely gas, which is at least partially ionized by both thermal processes and ionizing radiation from the white dwarf. Exterior to this point, the disk consists of relatively-large dust particles. These dust particles will be slightly charged due to collisions, but only up to a few electron charges. These properties have important consequences for the effects of magnetic fields on the pollution disk. 

There are two distinct magnetic truncation radii, at which the accreting material is bound to the magnetic field lines --- one for the gas and one for the dust. The gas magnetic truncation radius, $R_B^g$, takes the form \citep{Ghosh1978, Blandford1982}
\begin{equation}
    R_B^g=\omega\left(\frac{B_\star^4 R_\star^{12}}{GM_\star\dot{M}^2}\right)^{1/7}\,.
\end{equation}
In this equation, $B_\star$ is the stellar magnetic field strength, $M_\star$ is the stellar mass, $\dot{M}$ is the mass accretion rate onto the white dwarf, and $\omega\sim1$ is a scaling factor of order unity. 
This is the radius at which the pressure due to the magnetic field is balanced by the ram pressure of the accreting flow. Scaled to typical values, this equation becomes 
\begin{equation}\label{eqn:Rg}
\begin{split}
    \frac{R_B^g}{R_\sun}\simeq 3.84 &\left(\frac{B_\star}{\qty{e2}{kG}}\right)^{4/7}\left(\frac{R_\star}{\qty{0.01}{R_\sun}}\right)^{5/7}\\
    \times&\left(\frac{M_\star}{\unit{M_\sun}}\right)^{-1/7}\left(\frac{\dot{M}}{\qty{e10}{g/s}}\right)^{-2/7}\,.      
\end{split}
\end{equation}

However, this truncation radius only accounts for the strength of the magnetic field and does not address the issue of ionization, since neutral material will entirely ignore the magnetic fields. In contrast to the star formation problem, white dwarfs are small and dim enough that their magnetic fields can become significant exterior to where ionization occurs. There are two components of the disk, the dust and the gas, both of which may be ionized at different radii. 

We first compute a magnetic truncation radius for the slightly-charged dust disk $R_B^d$. This truncation radius is defined to be the point where the force on a dust particle due to the magnetic field is comparable to the force on the particle due to gravity. For a dust particle of mass $m_d$ at a distance $r$, the magnitude of the gravitational force is 
\begin{equation}
    |\boldsymbol{F}_G|=\frac{GM_\star m_d}{r^2}\,.
\end{equation}
If the dust particle carries a charge $q$, then the Lorentz force has a magnitude of 
\begin{equation}
    |\boldsymbol{F}_B|=qB\frac{v}{c}\,,
\end{equation}
where $v$ is the velocity of the dust particle with respect to the magnetic field. If the magnetic field is corotating with the white dwarf and the dust particle is in Keplerian motion, then $v=r(\Omega_\star-\Omega)$, where $\Omega_\star$ is the white dwarf rotation rate and $\Omega$ is the Keplerian rotation rate. Since $B(r)\simeq B_\star(r/R_\star)^{-3}$ and $\Omega=(GM_\star/r^3)^{1/2}$, equating $|\boldsymbol{F}_G|=|\boldsymbol{F}_B|$ means that the magnetic truncation radius of the dust disk is
\begin{equation}
    R_B^d=\left[\frac{\Omega_\star}{\sqrt{GM_\star}}+\frac{\sqrt{GM_\star}m_dc}{qB_\star R_\star^3}\right]^{-2/3}\,.
\end{equation}
The dust particles typically carry a few electron charges (e.g., \citealt{Ivlev2016, Balduin2023}). Even for dust masses as low as $m=\qty{e-9}{g}$ and magnetic field strengths of $B=\qty{e7}{G}$, the dust magnetic truncation radius $R_B^d<R_\star$, implying that the dust will never be coupled to the magnetic field lines. 

Therefore, the more critical point is where the gaseous disk is ionized. This depends heavily on the composition of the gas disk and the first ionization energy of its constituent elements. Alkali metals such as sodium and potassium will easily thermally ionize at the \qty{1500}{K} sublimation temperature and are dominant ion sources at this temperature (see, e.g., \citealt{Umebayashi1988, Balbus2000, Fromang2002}). Similar to the inner regions of protoplanetary disks, if these elements are present in any significant degree, then the resulting ionization fraction is $x_e\gtrsim\num{e-12}(n_n/\qty{e15}{cm^{-3}})^{-1/2}$, where $n_n$ is the neutral number density and $n_n\sim\qty{e15}{cm^{-3}}$ is typical for protoplanetary disks and for this circumstance (if the viscosity parameter $\alpha\simeq\num{e-4}$). At this ionization level and density, collisions between the ions and the neutrals will cause the gas to couple to the field. The ultraviolet and X-ray luminosity of the white dwarf will serve to further ionize this gas, supporting the coupling between the magnetic field lines and the disk. 

\begin{figure*}[t]
    \centering
    \includegraphics{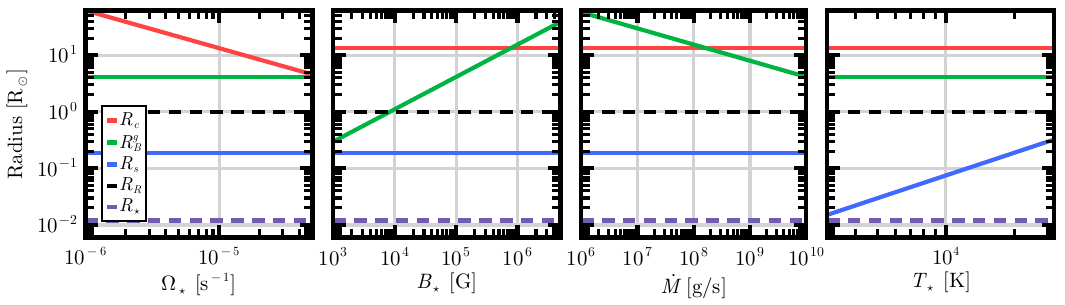}
    \caption{A comparison of the various important radii in the white dwarf accretion problem as a function of the stellar rotation rate $\Omega_\star$, magnetic field strength $B_\star$, mass accretion rate $\dot{M}$, and temperature $T_\star$. The relative location of these radii set the structure of the disk and the magnetic accretion flow. The Roche radius $R_R$ sets the outer edge of the debris disk, the corotation radius $R_c$ sets where magnetic coupling leads to accretion, the gas magnetic accretion radius $R_B^g$ sets where the gas could be accreted, and the sublimation radius $R_s$ sets where the dust is sublimated. The dust magnetic accretion radius $R_B^d$ is always less than the white dwarf radius, and so is not shown. }
    \label{fig:radiicomp}
\end{figure*}

In addition, even if sodium and potassium are not present in meaningful abundances, cosmic rays will ionize the tenuous surface of the disk.
The total disk surface density is approximately \qty{0.1}{g/cm^2}, significantly less than the $\sim\qty{100}{g/cm^2}$ attenuation length of the cosmic ray ionization rate \citep{Umebayashi1988}, so the ionization rate due to cosmic rays will be $\zeta\sim\qty{e-17}{s^-1}$ throughout the disk \citep{Spitzer1968}. 
At a temperature of $T=\qty{1500}{K}$ and $n_n\sim\qty{e15}{cm^{-3}}$, the corresponding steady-state ionization fraction is $\sim\num{4e-13}$, which is sufficient to couple the ions and neutrals \citep{Armitage2020}. 
While chemical network models are required to calculate the ionization fraction in detail, this toy model indicates that the cosmic rays are likely sufficient to quickly couple the gas to the magnetic field of the white dwarf.
The ionized (and magnetized) gas particles will then experience a negative torque from the white dwarf's magnetic field, since dust sublimation occurs inside the corotation radius (see Fig. \ref{fig:radiicomp}).

\subsection{Ordering of Radii}

The structure of the accretion disk is determined by the Roche radius $R_R$, the sublimation radius $R_s$, the gas magnetic accretion radius $R_B^g$, the dust magnetic accretion radius $R_B^d$, and the corotation radius $R_c$.
The ordering of these radii determines the structure of the pollution disk and the pattern of pollution on the surface of the white dwarf. 
The dominant factor is where the infalling material is coupled to the magnetic fields, at which point it will be accreted onto the surface of the white dwarf in a constricted region. 
In order for material to be coupled to the magnetic field, it must be within its corresponding magnetic accretion radius. 

Fig. \ref{fig:radiicomp} explores the structure of the pollution disk for a \qty{0.6}{M_\odot} white dwarf across a range of parameters, including magnetic field strength, rotation rate, mass accretion rate, and white dwarf temperature. 
We consider magnetic fields $B=\qtyrange{e3}{e7}{G}$, rotation rates $\Omega_\star=\qtyrange{e-6}{e-4}{s^{-1}}$, mass accretion rates $\dot{M}=\qtyrange{e6}{e10}{g/s}$, and temperatures $T_\star=\qtyrange{3000}{e5}{K}$. 

If the dust accretion radius $R_B^d$ is greater than the sublimation radius $R_s$, then the disk truncation radius $R_X=R_B^d$. Here, $R_X$ always represents the radius of the inner edge of the disk. If $R_B^d<R_s$, then the dust will not be coupled to the field at any point. In this case, the truncation radius $R_X$ is the minimum of the gas accretion radius and the sublimation radius. 
Over this parameter space, the gas accretion radius $R_B^g>R_s$. Since the dust magnetic accretion radius $R_B^d$ is always less than $R_\star$, the dust disk is never coupled to the magnetic field. Therefore, even for white dwarfs with somewhat weak magnetic fields, the accreting material will become truncated by the field approximately where the gas sublimates, since it will almost immediately become ionized. 

Note that if the corotation radius $R_c<R_X$, then the magnetic fields will act to accelerate the material and prevent pollution, rather than concentrating it onto the surface. In addition, if $R_X<R_\star$, the gas disk will process material directly onto the surface of the white dwarf and the pollution will be confined to a narrow band near the equator. 

\section{Accretion Beam Spatial Distribution}\label{sec:spatialdist}

The accreting material will generally be coupled to the magnetic field as soon as it becomes gaseous as ionization takes hold. This will have important consequences for the spatial distribution of metal pollution in the white dwarf's atmosphere. Once bound to the magnetic field lines, material accreted at a disk radius $R_X$ will be concentrated onto the stellar surface by the magnetic field. This section derives the size of the incoming pollution beam as it strikes the white dwarf's atmosphere.

\begin{figure}
    \centering
    \includegraphics{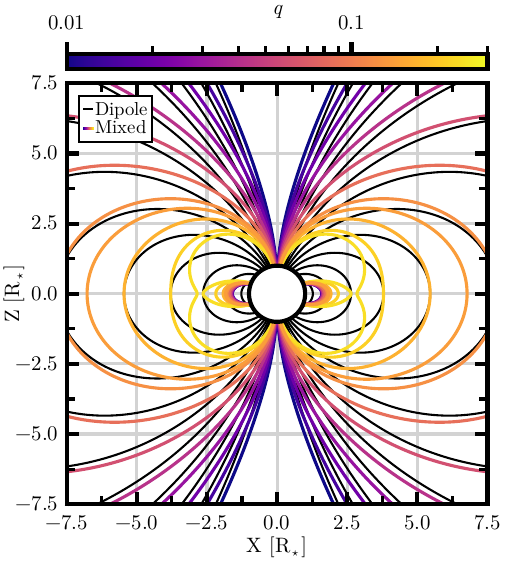}
    \caption{The two-dimensional spatial distribution of the magnetic field lines of the white dwarf. The dipole field is shown in black, while a mixed dipole-octupole field with $\Gamma=10$ is shown as colored lines. The color of the mixed field indicates the $q$ value that a field line corresponds to. Note that the small lobes of field at the equator share $q$ values with the larger field lobes. }
    \label{fig:Bfield}
\end{figure}

The magnetic field takes the form
\begin{equation}
\begin{split}
    \mathbf{B}=\frac{B_{\rm oct}}{2}\xi^{-5}&\Big[(5\cos^2\theta-3)\cos\theta\,\hat{r}\\
    &+\frac{3}{4}(5\cos^2\theta-1)\sin\theta\,\hat{\theta}\Big]\\
    +\frac{B_{\rm dip}}{2}\xi^{-3}&\left(2\cos\theta\,\hat{r}+\sin\theta\,\hat{\theta}\right)+\mathbf{B}_n\,. \label{eq:Bfield}
\end{split}
\end{equation}
In Eq. \eqref{eq:Bfield}, $\xi=r/R_\star$ is the dimensionless radius, $B_{\rm oct}$ is the octupole field strength, $B_{\rm dip}$ is the dipole field strength, and $\mathbf{B}_n$ are the high-order field terms. The polar angle $\theta$ is measured from the spin pole of the white dwarf. If $\mathbf{B}_n$ is assumed to be small, \citet{Adams2012} showed that the value 
\begin{equation}
    q=\frac{\Gamma}{4}\xi^{-3}(5\cos^2\theta-1)\sin^2\theta+\xi^{-1}\sin^2\theta\,,
\end{equation}
where $\Gamma\equiv B_{\rm oct}/B_{\rm dip}$, is conserved along magnetic field lines. 
For a given point in space, there is a single value of $q$, which specifies the field line that passes through that point. 
This value can then be used to identify the location where this field line passes through the surface of the white dwarf. 

Fig. \ref{fig:Bfield} shows the field lines of the white dwarf for both a pure dipole ($\Gamma=0$) and a mixed dipole-octupole field ($\Gamma=10$). 
As expected, the dipole field dominates at high radius, so the accretion locations will be equivalent.
On the other hand, the field lines in the mixed case are more tightly spaced at the system pole. As a result, the mixed field will concentrate the metal pollution into a narrow region relative to the dipole case.

\subsection{Accretion from an Eccentric Disk}

In order to derive the pollution region, consider a general case where the accretion/debris disk is eccentric and inclined by some angle $I$ relative to the white dwarf's spin axis. Due to the eccentricity of the disk, the accretion flow will originate from a single point on the disk (the periapsis) and land at a single point on the surface of the white dwarf. As the white dwarf rotates, the magnetic field will sweep across this accretion point and the material will be accreted onto different points on the white dwarf's surface. In addition, the magnetic field pole will not necessarily be aligned with the rotation axis, but offset by some polar angle $\beta$. As the white dwarf rotates, the angle $\theta$ between the magnetic field pole and the disk periapsis will thus vary. 

If $\xi_X=R_X/R_\star$ and the material is accreted from a polar angle $\theta$ relative to the magnetic pole, the angle $\theta_0$ at which the accreted material strikes the white dwarf surface will be the solution to
\begin{equation}
\begin{split}
    \sin^2\theta&\left[\xi_X^{-1}+\frac{\Gamma}{4}\xi_X^{-3}\left(5\cos^2\theta-1\right)\right]\\
    =\sin^2\theta_0&\left[1+\frac{\Gamma}{4}(5\cos^2\theta_0-1)\right]\,. \label{eq:theta0cond}
\end{split}
\end{equation}
Depending on the value of $\xi_X$ and $\Gamma$, there may be multiple solutions to Eq. \eqref{eq:theta0cond}. This captures the fact that octupole field lines may pass through the white dwarf twice in a single quadrant (see Fig. \ref{fig:Bfield}). In this case, the solution branch that passes through the accretion point should be used. In general, the accretion point is relatively far from the white dwarf, and the outer solution will be used. However, occasionally the inner branch is more appropriate. 

\begin{figure}
    \centering
    \includegraphics{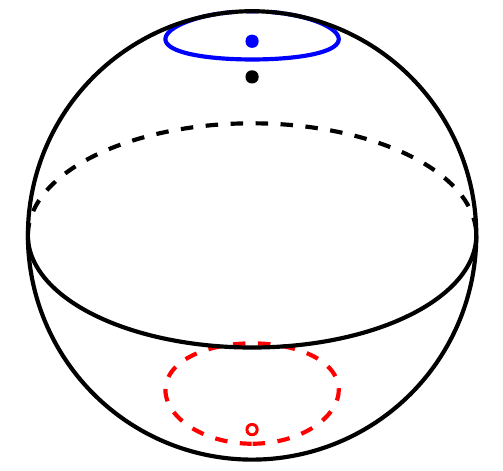}
    \caption{The accretion points on a white dwarf, with an angle $\beta=\pi/6$ between the magnetic pole and the spin axis, an inclination $I=\pi/4$ between the accretion disk and the spin axis, the octupole-dipole ratio $\Gamma=1$, and a disk truncation radius at $R_X=2R_\star$. The spin axis of the white dwarf is straight up, while the blue and red points are the poles of the magnetic field. The blue and red lines show where accreted metals will strike the atmosphere. For these parameters, the red line and point are on the far side of the white dwarf and are shown as dashed lines. The black line shows the magnetic field equator. Finally, the black point shows the direction where the material is accreted from. As the white dwarf rotates, the points where the accretion strikes the white dwarf's surface will sweep out the blue and red curves.}
    \label{fig:polmap}
\end{figure}

For a given magnetic field offset angle $\beta$ and a phase $\phi$ between the white dwarf's magnetic field pole and the accretion point,\footnote{The phase $\phi$ is defined such that when $\phi=0$ the spin pole, magnetic field pole, and accretion point are coplanar.} the spherical law of cosines provides the angle $\theta$ between the accretion point and the magnetic field pole, which can then be used to solve for $\theta_0$ in terms of $\beta$, $I$, and $\phi$.

For a given configuration of magnetic pole, spin pole, and accretion disk periapsis, the accretion flow will strike the white dwarf's surface at the two points that are (i) in the plane described by the magnetic pole and the accretion disk periapsis and (ii) are at an angle $\theta_0$ from the magnetic pole direction. 
Given these constraints, the position angles of the accretion point in the corotating frame of the white dwarf can be derived. 
The detailed equations for these values are given in Appendix \ref{appendix:accpt}.

As the white dwarf rotates, the value of $\phi$ will go from zero to $2\pi$ and the accretion point will move across the surface of the white dwarf. Fig. \ref{fig:polmap} shows the regions of the white dwarf where pollution will be concentrated for offset angle $\beta=\pi/6$, inclination $I=\pi/4$, octupole-dipole ratio $\Gamma=1$, and accretion distance $\xi_X=2$. The pollution will be symmetric across the plane normal to the magnetic field pole, since the accretion will spiral along the field lines in both directions and strike the surface twice (see Fig. \ref{fig:systemdiagram}). 

Once $\theta_0$ has been solved for, the width of the pollution region $\de\theta_0$ can be calculated. The chain rule provides that 
\begin{equation}\label{eq:dtheta0}
    \de\theta_0=\frac{\de R_X}{R_\star}\frac{\partial q}{\partial\xi}(\xi_X,I)\left[\frac{\partial q}{\partial \theta}(1,\theta_0)\right]^{-1}\,.
\end{equation}

Since the gas disk can be approximated as a Shakura-Sunyaev $\alpha$-disk, the accretion region in the disk will be approximately one scale height $H$ in width, since this is the length scale of the viscous spreading (see, e.g., \citealt{Hartmann2009}). The scale height at this point and the accretion region width is given by 
\begin{equation}
    \de R_X=H=\sqrt{\frac{2k_BT_sr^3}{GM_\star \bar{m}}}\,,
\end{equation}
where $T_s$ is the temperature of sublimation (\qty{1500}{K}) and $\bar{m}=\mu m_p$ is the average mass of the particle. This takes the scaled form 
\begin{equation}\label{eq:H}
    \frac{H}{R_\sun}\simeq10^{-4}\left(\frac{R_s}{\qty{0.1}{R_\sun}}\right)^{3/2}\!\!\left(\frac{M_\star}{\unit{M_\sun}}\right)^{-1/2}\!\!\left(\frac{\mu}{12}\right)^{-1/2}
\end{equation}
where $\mu$ is the average particle weight in atomic mass units. When combined with Eq. \eqref{eq:dtheta0}, this equation provides the width of the accretion band on the white dwarf surface.

\begin{figure}
    \centering
    \includegraphics{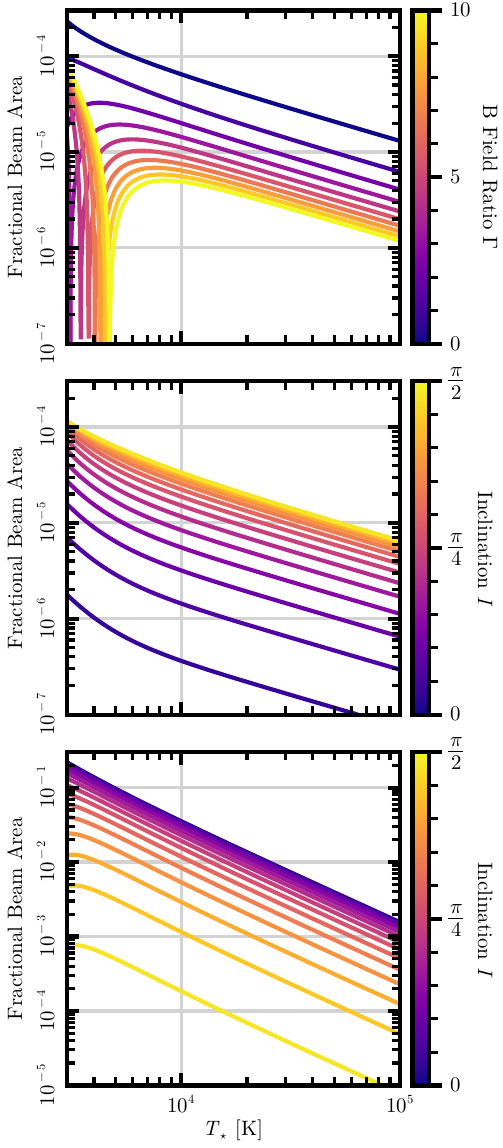}
    \caption{Fraction of the surface polluted by magnetic accretion as a function of the white dwarf temperature $T_\star$. The top panel shows the surface pollution fraction for different values of the octupole-dipole field ratio $\Gamma$ with the inclination of the disk fixed at $I=\pi/2$. The middle panel shows the pollution fraction for variable inclination with $\Gamma=1$. The bottom panel is the same as the middle panel, but the accretion disk is set to be circularly symmetric. In all cases, $\beta=0$. Note that the y-axis scale is different between the top two panels and the bottom panel. }
    \label{fig:coverfrac}
\end{figure}

Once accreted to the surface at $\theta_0$ with width $\de\theta_0$, the pollution beam will cover a total fractional area of $\sin\theta_0\de\theta_0$. This fractional pollution beam area versus stellar temperature, magnetic field parameter $\Gamma$, and accretion disk inclination $I$ is shown in Fig. \ref{fig:coverfrac}. In all cases, $\beta=0$ in this figure. As the white dwarf temperature increases, the sublimation radius $R_s$ increases, as does the scale height of the disk. This will therefore widen the accretion region. However, material accreted from larger radii will also strike closer to the pole of the white dwarf and concentrate the accreted material in a smaller area. It is clear from Fig. \ref{fig:coverfrac} that the concentration of material due to the magnetic field will dominate over the size of the accretion region, since the covering fraction scales negatively with temperature. 

In addition, the relative strength of the dipole and octupole fields has significant effects. Since the octupole component is more concentrated at the white dwarf's pole, stronger octupole fields (larger $\Gamma$ values) result in smaller pollution beams. In addition, for large $\Gamma$ values there is a critical radius at which the solution branch switches. This effect is the cause of the sharp drop in pollution beam areas around $T_\star\simeq\qtyrange{4000}{5000}{K}$ for large $\Gamma$ values. 

Finally, the pollution beam area increases as the material is accreted from closer to the magnetic equator and the inclination increases. Material from the equator is accreted onto relatively low latitudes, while material above the magnetic pole is only accreted onto the single point at the top of the white dwarf. Regardless of the value of these parameters, however, the covering fraction is never more than \num{e-4} and the pollution will only affect a small fraction of the white dwarf's surface. 

\subsection{Circularly Symmetric Accretion}

This calculation has assumed that the material is accreted from a single point on the disk, presumably the orbital periapsis. However, apsidal precession will cause the disk material's arguments of periapsis to diffuse away from this point and become uniform. Accretion from such a disk would then occur from a circular band. Note that even if each dust grain is on an eccentric orbit, if the arguments of periapsis are uniformly distributed then the magnetic accretion will be roughly rotationally symmetric in the disk plane. In this case, the accretion flow can be treated as coming from effective inclinations $[I,\pi/2]$ relative to the white dwarf spin/magnetic field axis. In such a circumstance, the fraction of the surface that is covered by the polluting beam is given by 
\begin{equation}
    f(I)=\cos(\theta_0(\xi_X,\pi/2))-\cos(\theta_0(\xi_X,I))\,.
\end{equation}
The polluting beam fraction for a white dwarf accreting from an inclined disk is shown in the bottom panel of Fig. \ref{fig:coverfrac}. For nearly circumpolar disks ($I\simeq0$), a significant fraction of the white dwarf's surface can be covered by the beam of accreting metals, although disks aligned with the white dwarf's rotation axis will necessarily be rare due to both spin-orbit coupling and simple geometric arguments.

\section{Vertical Sinking and Horizontal Spreading}\label{sec:diff}

\begin{figure*}
    \centering
    \includegraphics{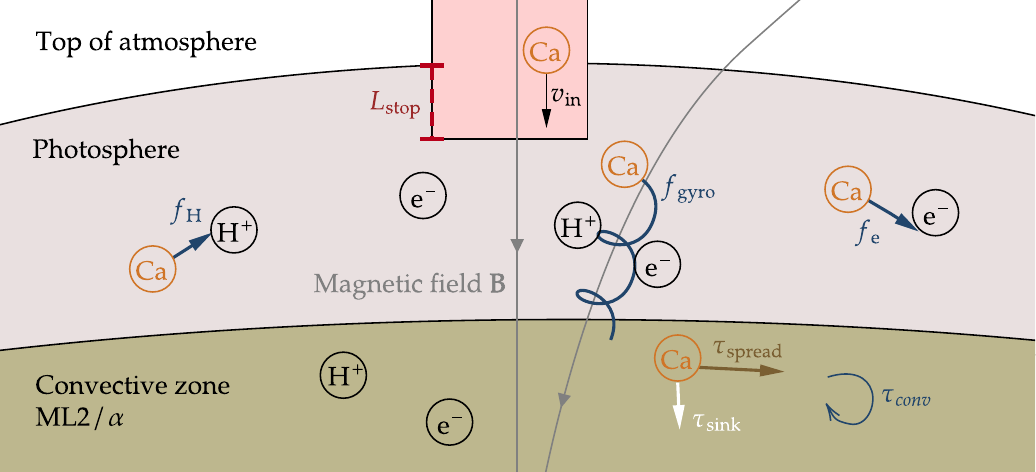}
    \caption{Diagram illustrating a DA (hydrogen-dominated) white dwarf's atmospheric structure and dynamics of incoming polluting $^{40}$Ca ions. Initially, the incoming particle arrived at the top of the atmosphere with velocity $v_\mathrm{in}$, punching through the atmosphere with length $L_\mathrm{stop}$ which is a small fraction of the atmospheric pressure scale height. Within the atmosphere, the particle's motion can be affected by collisions with atmospheric electrons and ions (characterized by frequencies $f_e$ and $f_H$ for collisions with electrons and hydrogen, respectively) or cyclotron motion from the background (characterized by the gyrofrequency $f_\mathrm{gyro}$). There may be a convection zone, which may be beneath or within the photosphere. The bulk motion of particles in the atmosphere can be described by the spreading timescale $\tau_\mathrm{spread}$ and the sinking timescale $\tau_\mathrm{sink}$. When $\tau_\mathrm{sink} \gtrsim \tau_\mathrm{spread}$, the initial pollution spot can spread globally before sinking, but a pollution gradient remains across the surface. When $\tau_\mathrm{sink} \gg \tau_\mathrm{spread}$, the white dwarf atmosphere will be homogeneously mixed. $\tau_\mathrm{conv}$ is the vertical convective turnover timescale. }
    \label{fig:atmosphere_diagram}
\end{figure*}

After the accreted metals strike the surface of the white dwarf, interactions with the white dwarf's atmosphere will modify the area of the pollution region. 
This section studies the dynamics and timescales for a beam of ionized material to sink and spread in the atmosphere of magnetic white dwarfs, which depends on the interactions with atmospheric plasma as well as the relative vertical sinking and horizontal diffusion timescales. 
The behavior of a $^{40}$Ca particle will be used as an example, since white dwarfs are typically polluted with rock-forming species like calcium, magnesium, silicon, and iron \citep[e.g.,][]{Zuckerman2007, Xu2021, Putirka2021, BadenasAgusti2024}.

A schematic of the typical behavior of this test particle in the atmosphere of a white dwarf is shown in Fig. \ref{fig:atmosphere_diagram}. The typical mass and charge of $^{40}$Ca ions are used for these calculations to illustrate the order of magnitude of various effects.
For an ionized $^{40}$Ca ion in the temperature ranges of a white dwarf atmosphere ($T_\star=\qtyrange{7000}{35000}{K}$), the ion mass $m_Z \simeq \qty{40}{m_p}$ and the typical charge state is $Z\simeq 2$.
Although the analyses in this section is focused on a polluting $^{40}$Ca ion, it can be readily applied to other polluting elements.

\subsection{Initial Stopping Length}

As studied in the previous sections, the polar regions of a magnetic white dwarf will receive a beam of polluting ions.
The penetration depth of this beam into the white dwarf atmosphere is of critical importance to determining the sinking and convective timescales. 
If the beam reaches below the photosphere, then these metals would be rarely detected.
Since the beam's velocity is initially tangent to the magnetic field lines, the incoming metals will experience no Lorentz force from the magnetic field. 
Hence, only the effects of the atmospheric plasma will be important in stopping the incoming beam of ionized metals.

The incoming particle will fall at roughly the free-fall velocity, so the incoming velocity $v_\mathrm{in}$ is determined by energy conservation such that \citep{Hartmann2016, Pham2024a},
\begin{align}\label{eqn:vin}
    v_\mathrm{in} =& ~\sqrt{2G M_* \left(\frac{1}{R_*}-\frac{1}{R_{B}^g}\right)} \approx \sqrt{\frac{2G M_*}{R_*}} \nonumber\\ \simeq& ~4.8 \times 10^8~\mathrm{cm~s^{-1}} \left(\frac{M_*}{\qty{0.6}{M_\odot}}\right)^{1/2} \left(\frac{R_*}{\qty{0.01}{R_\odot}}\right)^{1/2}\,,
\end{align}
assuming that $R_B^g\gg R_\star$ (see Fig. \ref{fig:radiicomp}).
As these particles can achieve a high incoming velocity, they can punch through a portion of the atmosphere before being thermalized.

The stopping length, which is the characteristic distance over which the incoming particle dissipates its incoming kinetic energy, is given by $ L_\mathrm{stop} \sim {v_\mathrm{in}} / {\nu_\mathrm{slow}}$.
Here, $1 / \nu_\mathrm{slow}$ is the characteristic timescale on which the particle is slowed down by the atmospheric ions.
Appendix \ref{appendix:stoppinglength} shows that $\nu_\mathrm{slow}$ is dominated by interactions with electrons.
That is, interactions with electrons are much more effective at slowing down the incoming particle, compared to interactions with atmospheric hydrogen ions.
Since the magnetic field can do no work to these particles, it has no impact on the stopping length.
The stopping length due to interactions with electrons is a fraction of the atmosphere's pressure scale height and takes the form
\begin{align}
    \frac{L_\mathrm{stop, e}}{L_\mathrm{P}} \approx&~ \num{2.4e-1}\left(\frac{v_\mathrm{in}}{\qty{4.8e8}{cm/s}}\right) \times\nonumber\\
    &~\left(\frac{\nu_\mathrm{slow}}{\qty{2.0e5}{s^{-1}}}\right)^{-1} \left(\frac{T}{\qty{2e4}{K}}\right)^{-1} \times \nonumber\\
    &~\left(\frac{M_*}{\qty{0.6}{M_\odot}}\right) \left(\frac{R_*}{\qty{0.01}{R_\odot}}\right)^{-2}\,.
\end{align}
The pressure scale height $L_\mathrm{P}$ for a $M_*=\qty{0.6}{M_\odot}$ hydrogen-dominated atmosphere white dwarf at $T=\qty{20000}{K}$ is given by $L_\mathrm{P} \sim (k_B T R_*^2) / (G m_p M_*) \simeq \qty{e4}{cm}$.
The expression for $\nu_\mathrm{slow} = \nu_\mathrm{slow, e}$ is given in Eq. \eqref{eqn:nu_e}.

Since the white dwarf photosphere's height is several pressure scale heights, the incoming particle punches through a small portion of the photosphere before it is significantly slowed down.
After the stopping length, the particle is thermalized, with subsequent motion dominated by either collisions or the magnetic field as it continues to sink down through the rest of the photosphere and into the convection zone (if there is a convection zone).

\subsection{Collisions and the Magnetic Field}

Once thermalized, these polluting particles may either follow the magnetic field lines or be dominated by scattering from hydrogen ions and electrons. 
To understand which process dominates, we compare the magnetic gyrofrequency to the collision rates between $^{40}$Ca ions and the white dwarf atmosphere's particles.
The magnetic gyrofrequency is defined as
\begin{equation}
    f_\mathrm{gyro} = \frac{Z e B}{m_Z} = \qty{4.8e8}{s^{-1}} \left(\frac{B}{\qty{e6}{G}}\right),
\end{equation}
where $e$ is the elementary charge, $B$ is the surface magnetic field strength, and $Z$ and $m_Z$ are the charge state and mass of the polluting particle.

The collision frequency between an arbitrary polluting ion and the atmospheric electrons takes the form \citep{Huba2013}
\begin{align}
    f_\mathrm{e} =~&\qty{6.4e11}{s^{-1}} \left(\frac{n_e}{\qty{5e16}{cm^{-3}}}\right)  \times \nonumber\\
    &\left(\frac{T}{\qty{2e4}{K}}\right)^{-3/2} \left(\frac{\ln \Lambda_e}{10} \right)\,,
\end{align}
where $n_e$ is the electron number density, $T$ is the temperature, and $\ln \Lambda_e\sim10$ is the Coulomb logarithm for electron collisions.
In addition, the ion-ion collision frequency between a polluting element and the hydrogen ions in the atmosphere is given by 
\begin{align}\label{eqn:ion_ion}
    f_{H} =&~ \qty{6.7e9}{s^{-1}}\left(\frac{n_H}{\qty{5e16}{cm^{-3}}}\right) \left(\frac{m_Z}{m_\mathrm{Ca}}\right)^{-1/2} \times \nonumber\\
    &\left(\frac{T}{\qty{2e4}{K}}\right)^{-3/2} \left(\frac{Z_H}{1}\right)^2 \left(\frac{Z}{2}\right)^2 \left(\frac{\ln \Lambda_i}{10} \right)\,.
\end{align}

The number densities of electrons $n_e$ and hydrogen ions $n_H$ for a completely ionized atmosphere are estimated to be  $n_e \approx n_H = \rho/m_p$, where $\rho$ is the atmosphere's density and $m_p$ is the proton mass.
MESA simulations (see Sec. \ref{sec:MESA}) reveal that the photosphere has a density on the order of $\rho \sim \qty{e-7}{g\,cm^{-3}}$, which corresponds to number densities $n_e = n_H \sim \qty{5e16}{cm^{-3}}$.

Comparing these values, $f_e \gg f_H \gg f_\mathrm{gyro}$ in white dwarf atmospheres.
Collisions between the polluting ion and electrons and atmospheric ions are thus orders of magnitude more important than cyclotron motion induced by the magnetic field.
In addition to gyration, the magnetic field can also induce a mirror force through the conservation of magnetic moment, which is capable of reflecting the incoming particle out of the atmosphere (as seen in the Earth's ionosphere; \citealt{Chen2016}).
However, this effect relies on the conservation of magnetic moment and requires that the background magnetic field must be slowly changing over the gyration timescale.
Since the system is strongly collisional once the particle is within the atmosphere, interactions with atmospheric electrons and ions will deflect the particle to nearby magnetic field lines over a gyration period, violating the magnetic moment adiabatic invariant and preventing the mirror force in this regime.
Since collision is the dominant process, polluting ions are quickly thermalized with the atmosphere through many interactions with atmospheric electrons and ions.

\subsection{Convection and Magnetic Fields}\label{sec:convection_timescale}

\begin{figure}
    \centering
    \includegraphics{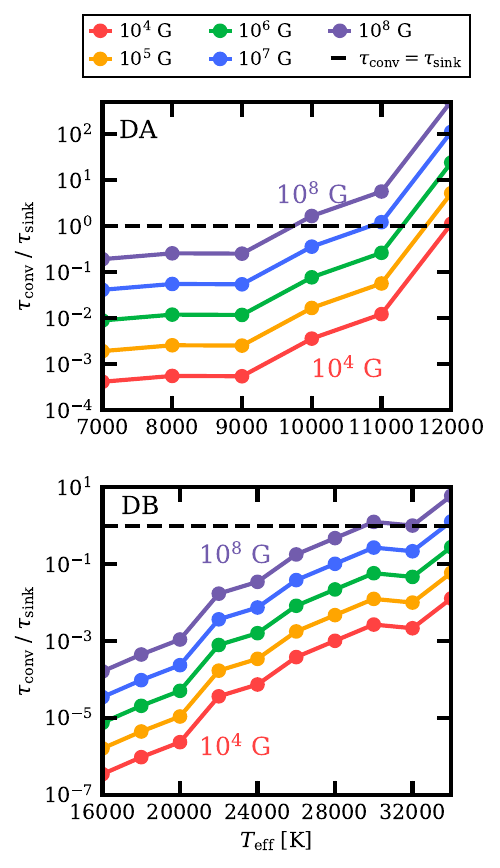}
    \caption{
    \textbf{Top}: The maximum convective turnover time to sinking time ratio, $\tau_\mathrm{conv}/\tau_\mathrm{sink}$, for a hydrogen-dominated atmosphere white dwarf (abbreviated as DA).
    The different colored curves show this ratio at various magnetic field strengths ranging from $B=10^4 - 10^8$ G, and the horizontal dashed line shows where the ratio is unity.
    When the ratio is much less than order unity, then the atmosphere is well-mixed.
    $\tau_\mathrm{conv}$ is calculated from Eq. \ref{eqn:tconv} with the atmosphere convective flux, scale height, and density calculated by MESA.
    The metal diffusion timescale at the bottom of the convection zone $\tau_\mathrm{sink}$ is measured by MESA.
    Quantities calculated in MESA assumed a regular convection zone with ML2/$\alpha=0.8$ \citep{Bauer2019}. 
    The rotation rate $\Omega$ is set by the rotation period $P$, which is chosen to be \qty{2}{hr}, consistent with many magnetic white dwarfs \citep[][]{Hernandez2024}.
    Note that for hydrogen-dominated atmosphere white dwarfs, when $T_\mathrm{eff} > \qty{12000}{K}$, the atmosphere becomes fully radiative. 
    \textbf{Bottom}: Same as top panel, except for a helium atmosphere white dwarf (abbreviated as DB), where the convection zone is simulated with ML2/$\alpha=1.25$ \citep{Koester2020}.
    }
    \label{fig:tconv_tsink}
\end{figure}

If the white dwarf's atmosphere is convective, then the sinking particle will encounter the convection zone (which may be beneath or even be contained within the photosphere), as seen in the diagram in Fig. \ref{fig:atmosphere_diagram}.
We now quantitatively study the effects of a magnetic field on convection.
By balancing buoyancy, Coriolis, and Lorentz forces, it can be shown that the (vertical) convective speed $v_B$ in the presence of magnetic field and rotation is (J. R. Fuentes 2025, private communication)
\begin{equation}\label{eqn:tconv}
    v_B \sim \left(\frac{F_\mathrm{conv}}{L_P}\right)^{2/3} \left(\frac{4\pi\rho}{2 \Omega B^2 L_P}\right)^{1/3}
\end{equation}
where $F_\mathrm{conv}$ is the convective flux, $L_P$ the pressure scale height, $\rho$ the density, $\Omega$ the rotation frequency, and $B$ the strength of the magnetic field. Note that $(F_\mathrm{conv} / L_P)^{1/3}$ is the convective speed in the absence of rotation and magnetic field.
For mixing length theory, the convective flux takes the form \citep[e.g.,][]{JoyceTayar2023}
\begin{equation}
    F_\mathrm{conv} = \frac{1}{2}\rho v c_P T \alpha\left(\nabla_T - \nabla_\mathrm{ad}\right)
\end{equation}
where $v, c_P, T, \alpha$ are the non-magnetic convective velocity, heat capacity, temperature, and mixing length parameter, respectively.
$\nabla_T$ and $\nabla_\mathrm{ad}$ are the temperature and adiabatic temperature gradients.

The convective turnover timescale $\tau_\mathrm{conv}$ takes the form 
\begin{equation}
    \tau_\mathrm{conv} \sim \frac{L_P}{v_B} \propto B^{2/3}.
\end{equation}
Since the turnover timescale scales with $B$ field strength, stronger magnetic fields will weaken convection.
In several $\tau_\mathrm{conv}$, the convection zone becomes well-mixed.
Note that since $\tau_\mathrm{conv}$ depends on atmospheric parameters, it varies across the convection zone.
Thus, we use the maximum value achieved by $\tau_\mathrm{conv}$, which is near the base of the convection zone, as the convective turnover timescale.
Finally, the turnover timescale is different than the sinking timescale, $\tau_\mathrm{sink}$, which is the timescale on which metals leave the convection zone and can no longer be observable.

In Fig. \ref{fig:tconv_tsink}, the ratio $\tau_\mathrm{conv} / \tau_\mathrm{sink}$ is shown for both hydrogen and helium atmosphere white dwarfs (DA and DB, respectively), at various magnetic field strengths.
Atmospheric parameters (density, heat capacity, temperature, gradients, etc.) and the sinking timescales are computed in MESA in a fully convective atmosphere (described in Sec. \ref{sec:MESA}).
For this calculation, we choose the white dwarf rotation period to be 2 hours, which is roughly (order of magnitude) consistent with magnetic white dwarfs' rotation period \citep[see table 2 and fig. 4 in][]{Hernandez2024}.
As seen, $\tau_\mathrm{conv} \ll \tau_\mathrm{sink}$ for most white dwarf cases, i.e., the metals are well-mixed throughout the convective zone.
This is in agreement to previous works studying the interplay between magnetism, convection, and element mixing \citep[e.g.,][]{Tremblay2015a, Cunningham2019}.

However, the morphology of the convection zone can still be significantly affected because $\tau_\mathrm{conv}$ can still be changed by orders of magnitude due to the magnetic field.
Therefore, the magnetic field may still impact the effective metal diffusion (sinking) timescale at the base of the convection zone, for fields in excess of $\sim\qty{100}{MG}$ (see Fig.\,\ref{fig:tconv_tsink}).
That being said, the effects of magnetism on the base of the convection zone is complex, and cannot be modeled with mixing length theory in 1D simulations.
Since our simulations are not capable of resolving the effects of magnetic fields on convection, sinking, or potential small-scale instabilities, this discussion is deferred to future work. 
Full 3D simulations of the atmospheres of magnetic white dwarfs are necessary to quantify the impact of the magnetic fields on the horizontal and vertical diffusion of the polluting metals.
As a first approximation, we consider the sinking timescale at the two extremes --- a complete convection zone and no convection zone. 
These limits represent where the magnetic field is least and most effective at impacting the convection zone.

\subsection{MESA and Sinking Timescale}\label{sec:MESA}

\begin{figure*}
    \centering
    \includegraphics[width=\linewidth]{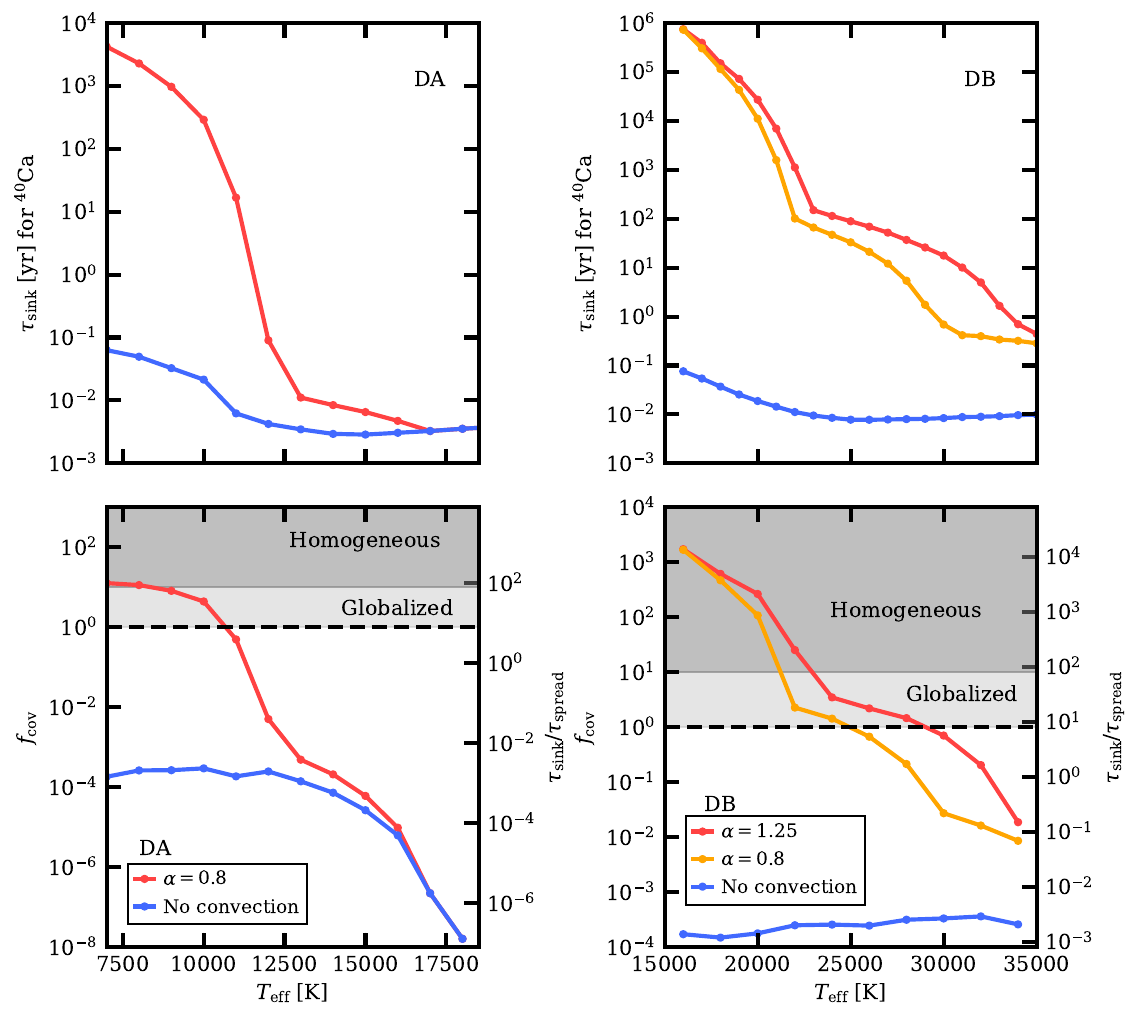}
    \caption{
    \textbf{Top left}: Sinking timescales of $^{40}$Ca on a hydrogen-atmosphere white dwarf (\qty{0.6}{M_\odot}, $\log g = 8$, abbreviated as DA) calculated by MESA. Two cases are simulated --- full convection with ML2/$\alpha = 0.8$ and no convection. This latter case represents the limit where the magnetic field significantly affects the base of the convection zone. The $\alpha=0.8$ curve corresponds to the fiducial mixing length value used in the MESA prescription of \citet{Bauer2019}.
    \textbf{Bottom left}: The pollution gradient parameter $f_{\rm cov}$ (Eq. \ref{eqn:coverage}) or the timescale ratio $\tau_\mathrm{sink}/\tau_\mathrm{spread}$. The horizontal diffusion coefficient $D_\mathrm{surf}$ is taken from the 3D radiation hydrodynamics code $\texttt{CO}^5\texttt{BOLD}$ \citep{Cunningham2021}, which is then used to calculate $\tau_\mathrm{spread}$ with Eq. \eqref{eqn:tspread}. When $f_\mathrm{cov}$ is of order unity, the pollution spot can spread over the entire white dwarf surface, but there is still a strong pollution gradient. Homogeneous pollution can be possible when $f_\mathrm{cov} \ll 1$. The two shaded regions with their corresponding labels show these two global and homogeneous pollution limits. Partial convection may be possible (Sec. \ref{sec:convection_timescale}), resulting in covering fractions between the convective and nonconvective limits shown in this figure. 
    \textbf{Right}: Same as the left panels, except now calculated for a helium-atmosphere white dwarf ($\log g = 8$, abbreviated as DB). Three cases are simulated --- convection with ML2/$\alpha=1.25$ \citep{Koester2020}, convection with ML2/$\alpha=0.8$ \citep{Cukanovaite2019}, and no convection. Note that the vertical scales differ for all four panels.  } 
    \label{fig:horizontal_spreading}
\end{figure*}

To calculate the sinking timescales in white dwarfs, we adopt the MESA \citep[version 10398,][]{Paxton2013} convection prescription by \citet{Bauer2019}, which will be summarized below.

The ML2 convection theory is used, following previous works \citep{Bohm1971, Koester2009, Koester2010}.
The fiducial mixing length parameter $\alpha$ has a value of $\alpha = 0.8$ \citep[as used in e.g.,][]{Tremblay2015}.
When the effective temperature $T_\mathrm{eff} > \qty{9000}{K}$, the gray iterative atmosphere in MESA gives appropriate sinking timescale when compared to previous results \citep{Koester2009, Koester2010}.
When $T_\mathrm{eff}$ reaches below \qty{9000}{K}, the atmosphere is switched to using pre-computed atmosphere white dwarf tables from \citet{Rohrmann2012}.
This switching prescription is only valid for white dwarfs with deep convection zones (where the optical depth exceeds $\tau_\mathrm{R} > 25$).
These simulations respect this condition, and in no-convection simulations, no switching is done.
The diffusion timescale of a trace element is calculated from MESA by injecting the metal at a rate of $\dot{M}=\qty{e7}{g/s}$.
When the system has reached equilibrium, the diffusion timescale is calculated according to \citep{Vauclair1979, Dupuis1992, Bauer2019}
\begin{equation}\label{eqn:sinking_timescale}
    \tau_\mathrm{sink} \equiv \frac{M_\mathrm{cvz}}{4 \pi r^2 \rho v_\mathrm{diff}},
\end{equation}
where the convection zone mass $M_\mathrm{cvz}$, radius $r$, density $\rho$, and diffusion velocity $v_\mathrm{diff}$ are all provided from MESA simulations.
$v_\mathrm{diff}$ is the sedimentation velocity of a polluting ion measured at the base of convection zone.
If there is no convection zone, the mass of the photosphere are taken instead.
Further details on this numerical setup can be found in \citet{Bauer2019}. 

In this section, we perform one set of MESA simulations with $\alpha = 0.8$, following previous work \citep{Koester2009, Koester2010, Bauer2019}.
Another set of simulations without convection are run to simulate when the magnetic field is most efficient at disrupting the metal diffusion timescale at the base of the convection zone.
Although partial impacts of this effect is possible, 1D MESA simulations cannot relate a given value of $\alpha$ to a specific magnetic field strength.
Therefore, we will only show the results for the two limiting cases.
Furthermore, while these models only simulate the diffusion of $^{40}$Ca, this methodology is reproducible with other polluting metals (as done in \citealt{Bauer2019}). 
We run models for both hydrogen-dominated (DA) and helium-dominated (DB) white dwarfs.
This section focuses on DA white dwarfs, while results for DB white dwarfs are discussed in Sec. \ref{subsec:DBWDs}. 

Fig. \ref{fig:horizontal_spreading} shows the sinking timescale as calculated by MESA simulations for a \qty{0.6}{M_\odot} white dwarf ($\log g = 8$).
As seen, atmospheres become fully radiative for all $\alpha$ when $T_\mathrm{eff} > \qty{17000}{K}$.
For a purely radiative atmosphere, there is no convection zone and the diffusion timescale is calculated using the mass of the photosphere.
As the white dwarf cools, hydrogen begins to recombine at temperatures $T_\mathrm{eff} < \qty{13000}{K}$, causing a rapid increase in opacity and inducing significant convection \citep{Tremblay2013}.
The general behaviors of the $\alpha=0.8$ models can be compared with results in \citet{Bauer2019} and \citet{Koester2020}.
In the no-convection case, the lack of a convection zone leads to a significantly shorter sinking timescale. 

\subsection{Horizontal Spreading}

Following \citet{Cunningham2021}, the horizontal and vertical diffusion of the trace polluting metal are described by a two-component model --- a horizontal diffusion component described by the coefficient $D_\mathrm{surf}$ and an averaged vertical sinking drift described by the velocity $v_\mathrm{drift}$.
In equilibrium (steady-state accretion and metal diffusion), the model can be described by
\begin{equation}\label{eqn:diffusion_eqn}
    D_\mathrm{surf}\nabla^2 u = v_\mathrm{drift}\frac{\partial u}{\partial z}.
\end{equation}
where $u = u(x,y,z)$ is the concentration (per unit area) of the trace metal as a function of position and $D_\mathrm{surf}$ is the horizontal spreading diffusion coefficient.
This equation assumes that the atmospheric metal concentration is in a steady-state. While the accretion can be episodic, introducing time dependence is deferred to future work. We discuss some potential consequences of asynchronous accretion in Sec. \ref{sec:caveats}.

Over the sinking timescale, $\tau_\mathrm{sink}$, it can be shown that the metal pollution gradient becomes (see Appendix \ref{appendix:diffusion})
\begin{equation}\label{eq:diffeqsoltau}
    u(R_\star, \tau_\mathrm{sink})\propto(D_\mathrm{surf}\tau_\mathrm{sink})^{-1}\exp\left(-\frac{\tau_\mathrm{spread}}{\tau_\mathrm{sink}}\right)\,
\end{equation}
where $t_\mathrm{sink}$ can be measured numerically, and $t_\mathrm{spread}$ is the spreading timescale defined by
\begin{equation}\label{eqn:tspread}
    \tau_\mathrm{spread} \equiv \frac{R_*^2}{4 D_\mathrm{surf}}.
\end{equation}
Clearly, when $\tau_\mathrm{sink} / \tau_\mathrm{spread} \gtrsim 1$, there is a small but nonzero pollution gradient between the accretion spot and the antipodal point.
Solving the spreading equation for the fraction $f_{\rm cov}$ of the white dwarf's surface that is covered by the pollution (see Appendix \ref{appendix:diffusion} and Eq. \ref{eq:fcovspread}) finds that
\begin{equation}\label{eqn:coverage}
    f_\mathrm{cov} \simeq \frac{1}{8} \frac{\tau_\mathrm{sink}}{\tau_\mathrm{spread}}.
\end{equation}

We emphasize that $f_{\rm cov}$ is not technically a covering fraction, but instead tracks the pollution gradient across the surface. 
When $f_\mathrm{cov}\ll1$, then $f_{\rm cov}$ is the fraction of the white dwarf surface where the abundance of polluted metals is $1/e$ of the maximum value. 
When $f_\mathrm{cov} \sim 1$, the pollution spot will spread globally, but there will still be a strong Gaussian pollution gradient with a magnitude of $1/e$. 
The atmosphere can only be homogeneously mixed with polluting elements when $\tau_\mathrm{sink} / \tau_\mathrm{spread} \gg 1~(\mathrm{or~}f_\mathrm{cov} \gg 1)$, such that the exponential term in Eq. \eqref{eq:diffeqsoltau} asymptotically approaches unity.

At this stage, the only unknown variable is the horizontal diffusion coefficient, $D_\mathrm{surf}$.
We will use the numerical values of $D_\mathrm{surf}$ from table 2 of \citet{Cunningham2021}, which result from the 3D radiation hydrodynamics code $\texttt{CO}^{5}\texttt{BOLD}$ \citep{Freytag2012}.
Note that these diffusion coefficients are calculated from non-magnetic white dwarf simulations \citep{Cunningham2019}.
However, since collisions dominate over magnetic gyration, the diffusion coefficients may still be correct to the order of magnitude.
While the magnetic field may suppress the small-scale eddies that are critical to horizontal diffusion, this would reduce the value of $D_{\rm surf}$. The pollution regions would therefore take up a smaller fraction of the white dwarf's surface, with important observational consequences.
Further 3D simulations will be necessary to calculate $D_\mathrm{surf}$ on magnetic white dwarfs.

The ratio $\tau_\mathrm{sink}/\tau_\mathrm{spread}$ and $f_\mathrm{cov}$ (the quantity controlling the pollution gradient) are also shown in Fig. 
\ref{fig:horizontal_spreading}.
The sinking timescale $\tau_\mathrm{sink}$ is taken from the MESA simulations described in Sec. \ref{sec:MESA}.
Temperatures are limited at the upper end at $T_\mathrm{eff} = \qty{18000}{K}$, since $D_\mathrm{surf}$ values from \citet{Cunningham2021} are not available at higher temperatures.
The two limiting cases with full and no convection are shown, with the expectation that partial impacts on $\tau_\mathrm{sink}$ results would be within those two limits. 

With full convection ($\alpha = 0.8$), only hydrogen atmosphere white dwarfs with temperature $T_\mathrm{eff} \lesssim \qty{12000}{K}$ (cooling age $\simeq\qty{3e8}{yr}$) will wash out any patchy pollution.
Thus, if they are accreting metals along their magnetic field lines, all young hydrogen-dominated white dwarfs with effective temperatures higher than $\sim \qty{13000}{K}$ are expected to exhibit localized pollution.
We again emphasize that due to Eq. \eqref{eq:diffeqsoltau}, even if the spreading timescale is sufficiently long to induce a globalized pollution signature, there is still a gradient of pollution over the white dwarf surface, which may be detectable.
As seen in Fig. \ref{fig:horizontal_spreading}, hydrogen-dominated atmosphere white dwarfs with full convection can only exhibit homogeneous pollution at temperatures below \qty{8000}{K}.
For hydrogen atmosphere white dwarfs above this temperature, there is always a gradient of pollution, even in cases where there is still full convection.

\subsection{Helium Atmosphere White Dwarfs}\label{subsec:DBWDs}

In this section, the previous calculations are repeated for white dwarfs with a pure helium atmosphere. 

The initial beam is strongly slowed by atmospheric electrons.
This beam can only punch through an initial distance of $L_\mathrm{stop, e} \simeq \qty{60}{cm}$, which is approximately 1\% of the atmospheric pressure scale height.

Within the atmosphere, the polluting elements' motion will again be dominated by collisions instead of magnetically induced cyclotron motion, since $f_e \gg f_\mathrm{He} \gg f_\mathrm{gyro}$.
In addition, these metal ions will be quickly thermalized due to many interactions with atmospheric particles.

The sinking timescale of a $^{40}$Ca ion on a pure helium-dominated atmosphere white dwarf is also calculated from MESA models.
The prescription in this case is identical to the case of a hydrogen-dominated white dwarf, except that we do not switch to pre-computed atmosphere tables at any point.
In addition, the version of MESA used in this paper does not have an appropriate opacity tables for helium below \qty{16000}{K} \citep{Bauer2019}, setting a lower limit on the simulation temperature.
Simulations are again performed with and without convection, using the ML2 convection theory in cases with convection.
For helium atmosphere white dwarfs, the mixing length parameter is typically set to $\alpha=\num{1.25}$ \citep[e.g.,][]{Koester2020}.
However, \citet{Cukanovaite2019} argues that $\alpha=0.8$ should be used for more accurate convection zone sizes.
We thus simulate three cases: $\alpha=1.25$, $\alpha=0.8$, and no convection.\footnote{
The sinking timescales for the ML2/$\alpha=\num{1.25}$ case are consistent with previous published values \citep{Koester2020}, indicating that this prescription is performing as expected. }

The top right panel of Fig. \ref{fig:horizontal_spreading} presents the sinking timescale for $^{40}$Ca on a helium atmosphere white dwarf ($\log g = 8$).
Similar to the case of a hydrogen-dominated atmosphere white dwarf, in the absence of convection the sinking timescale is several orders of magnitude lower than convective cases cases.
$f_\mathrm{cov}$ is shown in the bottom panel of the same figure. 
For convective helium atmosphere white dwarfs, globalized pollution is necessary at temperatures lower than \qtyrange{30000}{26000}{K} (depending on $\alpha$), and homogeneous surface pollution becomes necessary at temperatures lower than \qtyrange{20000}{22000}{K}.
When the sinking timescale is fully affected by the magnetic field, the pollution zones will not be spread to the entire white dwarf for the temperatures considered.

\subsection{Surface Covering with Initial Beam Size}\label{sec:covering_with_beam}

The discussion of convective spreading has so far assumed that the incoming beam arrives at a single spot at $x=y=0$.
Instead, the initial polluting beam will have some nonzero area, which for the sake of calculation will be described as a Gaussian with standard deviation $\sigma_\mathrm{beam}$.
It can then be shown that the solution to the diffusion equation again takes the form of a Gaussian (see Appendix \ref{appendix:diffusion}).
The resulting Gaussian simply modifies the previously found covering fraction to include the initial beam size.
$f_\mathrm{cov}$ (the quantity controlling the pollution gradient) is then estimated as
\begin{equation}\label{eqn:fcov_beam}
    f_\mathrm{cov} \simeq \frac{1}{8}\frac{\tau_\mathrm{sink}}{\tau_\mathrm{spread}} + \frac{\sigma_\mathrm{beam}^2}{4R_*^2}.
\end{equation}
Thus, the surface covering fraction results in Fig. \ref{fig:horizontal_spreading} can be easily modified by an additional term $\sigma_\mathrm{beam}^2 / 4 R_*^2$ to account for any incoming beam size.
Since the polluted areas generally span orders of magnitude, the pollution area can generally be taken to be the larger of the beam area or the surface covering fraction due to spreading.
With Eq. \eqref{eqn:fcov_beam}, we can fully describe the extent of pollution spreading on a white dwarf, given the incoming beam size, and the sinking and spreading timescales.

\section{Observational Signatures}\label{sec:obs}

\begin{figure}
    \centering
    \includegraphics{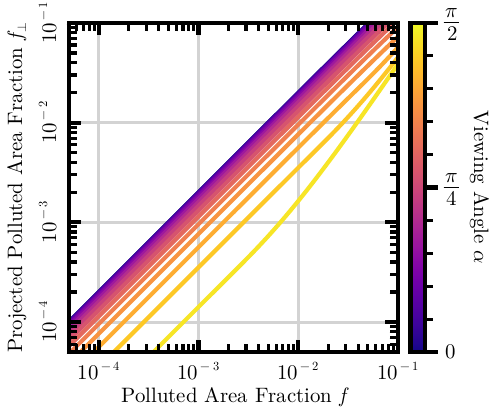}
    \caption{Polluted fraction of the projected visible surface for a given viewing angle and polluted fraction of the total surface. This calculation assumes that $\beta=0$ so that the projected surface is independent of the phase angle of the white dwarf's rotation. }
    \label{fig:projcoverfrac}
\end{figure}

\subsection{Projected Pollution Covering}\label{subsec:projfrac}

Previous sections compute the area of the white dwarf's surface that will be polluted by metals. However, observations of these objects do not sample the entire surface, but only a projected region. This section calculates the fraction of the projected area of the white dwarf that will be polluted in a given observation $f_\perp$. 

Consider a white dwarf where a fraction $f$ of the surface is polluted. Barring extreme disk or magnetic field geometries, these pollution regions will be concentrated into two bands near the magnetic poles of the white dwarf. These bands will be symmetric about the magnetic equator and close to axially symmetric about the pole. Any axial asymmetries will be smoothed out by the actions of horizontal convective diffusion. Since these regions will be primarily located at the poles, the convection will have the additional effect of concentrating the pollution towards the pole, so that the pollution region can be approximately modeled as two axisymmetric caps centered on the magnetic field poles. 

On the surface of the white dwarf, an area element $\de A(\theta,\phi)=R^2\sin\theta\,\de\theta\,\de\phi$ has a projected area $\cos\lambda\,\de A$, where $\lambda$ is the angle between the normal of the area element and the observation vector. If the observer's line of sight is inclined with respect to the magnetic field axis by a polar angle $\alpha$, the spherical law of cosines provides that the projected area element is 
\begin{equation} \label{eq:dAperp}
   \de A_\perp=R^2|\sin\theta\cos\phi\sin\alpha+\cos\theta\cos\alpha|\sin{\theta}\,\de\theta\,\de\phi\,.
\end{equation}
Each cap will extend to a polar angle $\theta_c=\cos^{-1}(1-f)$, where $f$ is the total fraction of the surface that is polluted. The total projected area of a single cap can thus be found by integrating Eq. \eqref{eq:dAperp} from $\theta=0$ to $\theta=\theta_c$.

Note that the observer only sees half of the surface of the white dwarf, and some of the pollution regions are hidden from sight. Since the pollution regions are azimuthally symmetric and mirrored across the equator, if any part of the upper region (say, at $\theta_0$, $\phi$) is in shadow, a corresponding portion in the lower region (at $\pi-\theta_0$, $\phi+\pi$) is visible. This can be seen in Eq. \eqref{eq:dAperp} by taking $\theta\rightarrow\pi-\theta$ and $\phi\rightarrow\pi+\phi$, which does not change $\de A_\perp$. As a result, only the area of a single cap is necessary. Therefore, the projected area is given by 
\begin{equation}\label{eq:fperp}
    f_\perp=\frac{1}{\pi}\!\!\int\displaylimits_0^{2\pi}\!\!\de\phi\!\!\int\displaylimits_0^{\theta_c}\!\!\de\theta|\sin\theta\cos\phi\sin\alpha+\cos\theta\cos\alpha|\sin{\theta}\,.
\end{equation}
This integral is numerically evaluated with the results versus $f$ and $\alpha$ shown in Fig. \ref{fig:projcoverfrac}. In general, $f_\perp$ and $f$ have the same order of magnitude, with slight corrections for the viewing angle $\alpha$. 

\subsection{Effects on Measured Mass Accretion Rate}

The concentration of the pollution into a fraction of the white dwarf's surface may have significant impacts on the measured accretion rate. 
Generally, the metal accretion rate onto a white dwarf is derived from the metal line strength in the atmosphere. 
A given line shape (e.g., equivalent width), in combination with the atmospheric parameters (e.g., temperature, pressure, composition), implies a certain metal abundance, which must be maintained by accreting material. 
However, the conversion from line shape to metal abundance generally assumes that the pollution is spread out across the entire surface (e.g., \citealt{Dupuis1993}). 

If the pollution is actually concentrated in some visible fraction $f_\perp$, then (to first order) the actual element abundance $X$ in the polluted region is related to the derived abundance $X_{\rm obs}$ by $X_{\rm obs}=f_\perp X$.
This equation assumes that the line strength and shape correlates linearly with both the abundance in the pollution region and the area of the pollution region. While there will be some nonlinear effects due to changes across the curve of growth, the metal content in the unpolluted white dwarf atmosphere, and limb darkening (see \citealt{Cunningham2021}), this relation is correct to order unity. 

If the element abundances are close to steady-state, then the calculated mass accretion rate of a given element will also be related to the actual mass accretion rate by a factor $f_\perp$ so that 
\begin{equation}\label{eqn:accretion_rate}
    \dot{M}_{\rm obs}=f_\perp\dot{M}\,.
\end{equation}
As we have shown in Secs. \ref{sec:accdisk} and \ref{subsec:projfrac}, $f_\perp\sim\numrange{e-4}{e-1}$, with significant dependence on the system geometry. The accretion rates measured for these polluted white dwarfs may thus be orders of magnitude lower than the true values. 

\subsection{Variability}\label{subsec:var}

If the magnetic fields concentrate the pollution to small regions of the white dwarf's surface, then the strength of the metal lines should vary over the rotation of the star. Here, the observation viewing angle $\alpha$ is defined to be with respect to the spin axis, $\beta$ is the angle between the spin axis and the magnetic field axis, and $\varphi$ is the phase angle of the rotation such that $\varphi=0$ when the magnetic field points towards the observer. Using the spherical law of cosines, the angle between the magnetic field pole and the observer, here defined to be $\lambda$, is 
\begin{equation}\label{eq:lambdavar}
    \cos\lambda=\sin\alpha\sin\beta\cos\varphi+\cos\alpha\cos\beta\,.
\end{equation}
Using $\lambda$ instead of $\alpha$ in Eq. \eqref{eq:fperp} then captures the dependence of the projected fraction on the phase angle.
Note that if $\beta=0$, then Eq. \eqref{eq:lambdavar} has no $\varphi$ dependence. That is to say, there will be no variability if the white dwarf's spin pole and rotation axis are aligned. 

\begin{figure}
    \centering
    \includegraphics{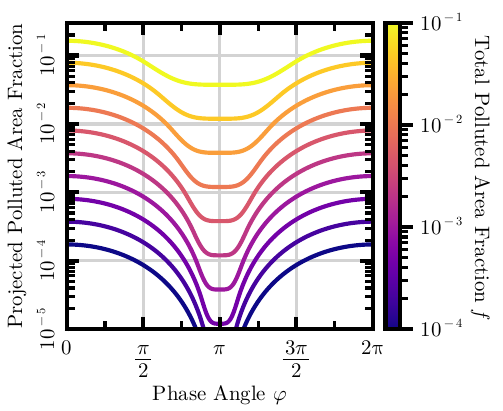}
    \caption{Variability in the projected cover fraction from the rotation of the white dwarf. The viewing angle is set to $\alpha=\qty{60}{\degree}$ and the magnetic field inclination is set to $\beta=\qty{15}{\degree}$, matching the model for WD 2138-332. The total polluted area fraction is set to vary from \numrange{e-4}{e-1}.}
    \label{fig:varrot}
\end{figure}

Fig. \ref{fig:varrot} shows the variability in the projected polluted area for a white dwarf with $\beta=\qty{15}{\degree}$, $\alpha=\qty{60}{\degree}$, and a range of total covering fractions. These values are chosen to reflect the parameters found for WD 2138-332, which exhibits clear variability in the metal line strength and magnetic field strength \citep{Bagnulo2024b}. The phases and periods of these variable signals are identical, strongly suggesting that magnetic accretion is responsible. 

If the metal line strength is proportional to the projected polluted area, then the \qty{0.5}{dex} variability in the line strength corresponds to \qty{0.5}{dex} in the projected area fraction. In Fig.~\ref{fig:varrot}, the topmost curve has a variability of $\sim\qty{0.5}{dex}$ and corresponds to a total polluted area of approximately \qty{10}{\percent}.
In addition, the variability curve calculated from this model is approximately sinusoidal for this $f$ value, matching the variability in the data. For white dwarfs with significantly smaller polluted areas, the variability in the metal lines will follow a non-sinusoidal function that may be detectable. 

Of course, the variability will be more complex than described here, with some additional noise that is difficult to parametrize. However, the calculation here provides an order-of-magnitude estimate for the variability that could be attributed to rotation. If aperiodic variations in measured mass accretion rate are observed on timescales different that the rotation period, then episodic accretion is likely responsible, rather than rotation. 

\section{Discussion}\label{sec:disc}

\subsection{Observed Variability and Spin Axis}

An important conclusion from Sec. \ref{subsec:var} is that variability in the projected pollution fraction over the white dwarf's spin (e.g., as detected by \citealt{Bagnulo2024a} and \citealt{Bagnulo2024b}) requires misalignment between the magnetic field pole and the spin pole of the system. If the field axis and the spin pole are aligned, then no such variability will be detected. This is particularly important in the context of white dwarf magnetic field measurements, which have found that the magnetic fields of white dwarfs are symmetric about their rotation axes \citep{Bagnulo2024d}. If the pollution is beamed onto the star by magnetic fields, then variability will not be detected in these systems. As a consequence, it is possible that significantly more polluted white dwarfs are only metal-enriched in a small region, since this patchy pollution could be obscured by the alignment of the spin and magnetic field poles. For these white dwarfs, the accretion rates measured via pollution abundances would be underestimated by orders of magnitude and difficult to detect. Note that measurements of the mass accretion rate via X-ray emission luminosities would not be biased in this manner.\footnote{While the one measurement of the mass accretion rate through X-ray luminosity is consistent with the spectroscopic measurements \citep{Cunningham2022}, further observations are needed to determine the impact of this mechanism.}

However, in circumstances where this variability can be detected, the magnitude of the variability and the shape of the phase dependence depend strongly on the polluted fraction of the white dwarf's surface. Measurements of the variability can thus constrain the pollution area and thereby jointly constrain the convective and accretion properties of the white dwarf. 

Finally, we note that not all photometric variability can be caused by an inhomogeneous metal distribution on a white dwarf surface. Indeed, a significant fraction of magnetic white dwarfs do not exhibit signatures of metal pollution \citep{Farihi2024}. This variability is presumably the result of some form of magnetic field--induced stellar activity, such as a spot, rather than heterogeneous metal pollution. 

\subsection{Observed Magnetic, Metal-Polluted White Dwarfs}

The two magnetic, polluted white dwarfs found so far with evidence of metal patches have temperatures $T_\star\simeq\qty{6250}{K}$ and $T_\star\simeq\qty{7150}{K}$ \citep{Bagnulo2024a, Bagnulo2024b}.
These white dwarfs also both host helium-dominated atmospheres.
This work shows that the inhomogeneous surface covering on these white dwarfs can be explained by the concentration of accretion from the debris disk.
We emphasize that the observations of \citet{Bagnulo2024a,Bagnulo2024b} and \citet{Farihi2024} themselves do not unambiguously demonstrate suppression of convection, nor do they directly measure spatial variations in the total mixed mass across the stellar surface; those works only detect rotationally modulated line and field variations, which we interpret here as one possible signature of localized accretion.
As shown in Fig. \ref{fig:coverfrac}, a small relative inclination between the polluting gas disk and the magnetic pole can cause a large surface covering fraction, up to 10\% of the white dwarf's surface.
Notice that this geometry requires either that the debris disk is circumpolar or that the magnetic field is highly inclined relative to the spin pole. 
The observational data suggest that either $\beta=\qty{15}{\degree}$ and $\alpha=\qty{60}{\degree}$ or that $\beta=\qty{60}{\degree}$ and $\alpha=\qty{15}{\degree}$, with uncertainty due to degeneracies \citep{Bagnulo2024b}. 
While magnetic field generation mechanisms would certainly prefer the $\beta=\qty{15}{\degree}$ scenario, a debris disk is far more likely to orbit the magnetic pole if $\beta=\qty{60}{\degree}$, leaving some degeneracy.
If this variable pollution is the result of abundance gradients, then either convection or horizontal spreading must be suppressed, perhaps by the magnetic field (see Sec. \ref{sec:convection_timescale} and Fig. \ref{fig:horizontal_spreading}).
If there is no such suppression, then the horizontal convection will easily cause the pollution regions to spread to the entire white dwarf surface, which is inconsistent with the observations.

On the other hand, it is also possible for the presence of a magnetic spot to alter the atmospheric structure such that the equivalent width of a given metal line varies over the white dwarf's surface, even if the composition is homogeneous. 
For example, a starspot produced by the threading of the magnetic field through the photosphere would have a different temperature than the rest of the photosphere. 
The metals in these spots would thus have a different equivalent width than the background photosphere.
Such a scenario would explain the periodically variable metal line strength while allowing for convection.
Future studies on line production in magnetic metal-polluted atmospheres will be necessary to constrain these possibilities. 

\subsection{Multiple Polluting Elements}

Fig. \ref{fig:horizontal_spreading} shows the surface covering fraction for polluting $^{40}$Ca ions.
However, white dwarfs are polluted with a combination of multiple elements including carbon, oxygen, silicon, and magnesium \citep{Xu2021}.
Furthermore, the two magnetic white dwarfs found in \citet{Bagnulo2024a, Bagnulo2024b} show pollution signs of magnesium and calcium among other heavy metals.
Although these elements may be expected to accrete within a single beam, the significantly different sinking timescales \citep{Koester2009, Koester2020} means that in this case these elements will pollute different areas of the white dwarf's surface, in concentric annuli (see Eq. \ref{eqn:coverage}).
Observationally, these concentric pollution regions mean that different elemental lines will exhibit different amplitude variability.
This effect can potentially explain the differences in variability amplitudes between elements in Fig. 2 of \citet{Bagnulo2024b}, since the longer diffusion timescale of magnesium versus calcium or iron \citep[e.g.,][assuming convection is unaltered]{Koester2020} is approximately the correct factor to explain the extra \qty{0.08}{dex} of variability in calcium and magnesium abundances. 

\subsection{Inferred Accretion Rate}

As shown in Fig. \ref{fig:radiicomp}, even weak magnetic field strengths as low as \qty{e3}{G} can cause metal patches on white dwarfs that cover fractions of a percent of the surface area (Figs. \ref{fig:coverfrac}, \ref{fig:horizontal_spreading}, \ref{fig:projcoverfrac}).
At the same time, Eq. \eqref{eqn:accretion_rate} shows that the observed surface covering fraction of the metal spot may cause inferred accretion rates to be orders of magnitude lower than the reality.

If we interpret the observed variability on WD 2138-332 \citep{Bagnulo2024b} as a signature of surface metal abundance variations, then under the model developed in this work, this variability is consistent with a pollution patch covering only \qty{10}{\percent} of the white dwarf surface.
This model could be tested if further detections of metal-line variability continue to show covering fractions of this order of magnitude.

In context, the current inferred white dwarf accretion rate are $\sim 10^6 - 10^{10}~\mathrm{g~s^{-1}}$, although these estimates are estimated from modeling of the photospheric metal lines and are dependent on accretion and stellar atmosphere models \cite[e.g.,][]{Farihi2012, Bauer2018, Koester2020, Brouwers2023, Blouin2022}.
X-ray observations provide an independent measurement of the accretion rate in the polluted white dwarf G29-38, where the accretion rate is found to be $\sim \qty{1.6e9}{g~s^{-1}}$.
Furthermore, G29-38 is a magnetic white dwarf, with a $3\sigma$ upper limit on the magnetic field strength estimated at $\sim 1.5~\mathrm{kG}$ \citep{Farihi2018, Cunningham2022}.
Future X-ray observations of weakly magnetic polluted white dwarfs may offer an opportunity to provide an independent constraint on the higher intrinsic accretion rates predicted by the model presented in this work.

Finally, a potential increase of up to \qty{3}{dex} in inferred accretion rate may require volatiles in the accretion disk to facilitate accretion rates higher than can be driven by the Poynting-Robertson effect \citep{Okuya2023}.
In turn, this implies that volatile rich reservoirs\footnote{Volatile processing in the accretion disk can produce rocky-like composition on white dwarf atmospheres \citep{Okuya2023}. In addition, the magnetic field can also prevent volatiles from accreting, likewise producing observed rocky-like polluted white dwarf compositions \citep{Zhou2024}.} may be required to produce observed pollution \citep[e.g.,][]{OConnor2023, Akiba2024, Pham2024}.
At the same time, adding up to \qty{3}{dex} to measured accretion rates may pose challenges for some proposed dynamical mechanisms to deliver materials to white dwarfs.
Mechanisms relying on small bodies reservoirs like exo--Kuiper Belts, scattered disks, and Oort Clouds \citep[e.g.,][]{Smallwood2018, OConnor2022, OConnor2023, Akiba2024, Pham2024} will be most affected, since those reservoirs must be also orders of magnitude more massive to explain the accretion rates.
This may place the necessary masses in conflict with existing estimates and constraints from solar system theory and observations.

\subsection{Magnetic and Polluted White Dwarf Population}

Directly resolving the polluted surface covering fraction for the existing population of polluted white dwarfs is a challenging endeavor for two reasons.
First, the magnetic field strength must be measured (e.g., by Zeeman splitting), which can be difficult to constrain due to low spectral resolution, especially with cooler white dwarfs.
Second, spectroscopic variability \citep[similar to those in][]{Bagnulo2024b} would be needed to constrain the metal spot size.
If the magnetic field is undetected or if the field pole aligns with the spin pole, then no variability will be detected and the pollution concentration may be orders of magnitude larger than expected.

These two challenges are exacerbated for the population of old and strongly magnetic white dwarfs.
One proposed mechanism to produce strong magnetic fields on white dwarfs is convective dynamo caused by crystallization \citep{Isern2017, Fuentes2024}.
This process can only begin to become effective for old white dwarfs with cooling ages of a few \unit{Gyr} \citep{Ginzburg2022}.
At the same time, if metal settling or horizontal mixing is inhibited (Sec. \ref{sec:convection_timescale}), then the metal spot can be orders of magnitude smaller than the surface area (Fig. \ref{fig:horizontal_spreading}).
Thus, these old, magnetic, polluted white dwarfs are much more likely to have a small metal pollution spot that could affect their inferred accretion by up to several orders of magnitude.
Simultaneously, these objects are relatively faint, thereby reducing the chance of constraining the pollution spot size to correct for the inferred accretion rate.

Since direct detection of pollution spots is challenging, one may instead approach the problem as a population.
The two statistical questions are: (i) what fraction of all white dwarfs are both polluted and magnetic and (ii) what fraction of polluted white dwarfs are magnetic?
Note that if the fraction of polluted white dwarfs with magnetic fields and the overall fraction of white dwarfs that are polluted is known, the product of the two is the fraction of all white dwarfs that are both polluted and magnetic. 

One approach to answer these questions is through the observed magnetic and polluted population in volume-controlled surveys, as done in \citet{OBrien2024} for the local 40 pc volume.
Out of the $1076$ white dwarfs identified in this work, 17 are magnetic and polluted (12 helium and 5 hydrogen), giving an occurrence rate of $\sim 2\%$ for the entire white dwarf population.
This subset of the white dwarf population is the case of interest for this paper.
However, \citet{OBrien2024} found that only 11\% of their sample are metal polluted, as characterized through optical spectroscopy, a much lower rate than previously found with UV spectroscopy \citep[25-50\%, e.g.,][]{Koester2014, OuldRouis2024}.
Thus, a larger population of weakly magnetic and weakly polluted white dwarfs may remain undetected in this survey.

Another approach is to independently assess the frequencies for magnetism and pollution.
This approach is motivated by the assumption that mechanisms for generating magnetic fields in white dwarfs are independent of the dynamical processes responsible for white dwarf pollution.
\citet{Bagnulo2021} performed a volume-limited survey of magnetic white dwarfs in the local 20 pc volume, with sensitivity $B > \qty{1}{kG}$.
This field strength is the lower limit of the range considered in this work. 
The authors found that $22 \pm 4 \%$ of white dwarfs are magnetic, consistent with the magnetism occurrence rate in \citet{OBrien2024}.
In parallel, the occurrence rate of white dwarf pollution in UV spectroscopy is 25-50\%  \citep{Koester2014}, with recent work revising the rate to 45\% \citep{OuldRouis2024}.
Thus, the probabilities for polluted and magnetic are $P(\mathrm{polluted}) = 0.25-0.5$ and $P(\mathrm{magnetic}) = 0.22$, respectively.
Under the assumption magnetism and pollution are independent events, we expect that about 
$P(\mathrm{polluted} ~\cap~ \mathrm{magnetic}) = P(\mathrm{polluted}) P(\mathrm{magnetic}) \sim 5 - 10\%$ of all white dwarfs should be detectable as both magnetic and polluted when harnessing UV spectroscopy, or 1--2\% when limited to optical spectroscopy only.

Furthermore, we can estimate the magnetism occurrence rate for the known polluted population, which is the conditional probability $P(\mathrm{magnetic} | \mathrm{polluted}) = P(\mathrm{magnetic}~\cap~ \mathrm{polluted})~/~P(\mathrm{polluted}) \simeq 0.2$.
That is, we currently expect that one in five known polluted white dwarfs should have detectable magnetic fields and could be subject to the results of this paper.
Therefore, follow-up multi-epoch time-series spectroscopy on the known polluted population may be able to uncover more localized pollution patches like on WD 2138-332.

The frequency of magnetism is not uniform across the entire white dwarf population, but correlates with white dwarf age and mass.
This dependence is highlighted by the properties of two clear subpopulations of magnetic white dwarfs --- young, massive magnetic white dwarfs versus old, low-mass magnetic white dwarfs \citep{Bagnulo2022, OBrien2024, Moss2025}.
While white dwarfs with magnetic fields strong enough to cause pollution spots ($B \gtrsim \qty{1}{kG}$) may be common, their occurrence rate is complicated by correlation with age, mass, and observation biases.
The effects described in this paper are expected to be more significant for older or more massive white dwarfs, and could be systematically affecting the inferred accretion rate for polluted white dwarfs in those subpopulations.

\subsection{Caveats}\label{sec:caveats}

The models and results present herein have several important caveats.

First, we have assumed that the magnetic fields of the white dwarf can be described as a mixture of a dipole and octupole component, especially far from the surface. 
Since each higher-order term decays more rapidly than the previous term, this assumption will be valid far from the surface of the white dwarf.
Near the surface, higher-order terms may dominate and divert the material from the expected dipolar position. 
However, this effect would likely induce changes in the pollution beam area of order unity, rather than orders of magnitude.
Regardless, the coupling of the material to the field lines means that material will continue to be concentrated in regions of high magnetic flux. 
Other factors such as diamagnetic screening \citep{Cumming2001} may systematically lower the magnetic field strength in accreting systems. 

Second, if small dust particles suppress the ionization fraction of the gas, then the gas may not be well-coupled to the magnetic field and will not be accreted along the field lines.
In this instance, the material will accrete to the white dwarf's equator and likely continue to cover a small area of the white dwarf's surface. 
This scenario can also occur if ionization is low due to high gas disk densities or low Na/K abundances. Further work is necessary to understand the precise conditions for magnetospheric accretion in these systems.

Third, our results cannot be directly correlated to an exact magnetic field strength, since we can only perform MESA simulations for cases with and without convection.
These simulation results are limited by available horizontal spreading data and opacity tables.
Furthermore, as mentioned in Sec. \ref{sec:convection_timescale}, magnetic fields may induce small-scale instabilities, and many other complex effects on the sinking and spreading timescales.
These effects cannot be captured by 1D MESA models.
To this end, we call for future works to simulate a full 3D atmosphere in the presence of a magnetic field, with various strengths at various temperatures.
This is particularly important for cooler white dwarfs such as those with detected patchy accretion, especially with emerging evidence pointing to cooler white dwarfs having stronger magnetic fields \citep{Camisassa2024, Moss2025}.

Fourth, this work uses a horizontal spreading timescale that is calculated for a non-magnetic white dwarf from \citet{Cunningham2021} and may not apply in the presence of a strong magnetic field.
Unfortunately, this is the only such timescale calculation that exists in the literature.
That being said, since collisions with atmospheric ions and not magnetically induced motion dominate the dynamics of the metal ions, the horizontal spreading timescale may be unaffected by strong magnetic fields.
If the magnetic fields are important, they will only suppress the horizontal convection and thereby shrink the polluted area. As we have shown, this would imply more significant underestimates of the mass accretion rate.

Fifth, thermohaline mixing has been predicted to affect the sinking timescales and inferred accretion rates by orders of magnitude \citep{Bauer2018, Bauer2019}.
Specifically, including this effect leads to accretion rates \qtyrange{3}{4}{dex} higher than previously inferred in hot DA white dwarfs ($T_\mathrm{eff} >\qty{15000}{K}$).
Likewise, convective overshoot has been predicted to affect the sinking timescale in hydrogen atmosphere white dwarfs by \numrange{1}{2} orders of magnitude \citep{Cunningham2019}, with an accretion rate up to \qty{1}{dex} higher than generally inferred.
While neither of these effects are implemented in our simulations, our results are reported in terms of the ratio $\tau_\mathrm{sink}/\tau_\mathrm{spread}$ and thus can be readily used with any updated sinking timescales.

Sixth, the polluting particle is assumed to be a trace component of the atmosphere.
That is, this calculation has assumed that the number density of metals $n_Z \ll n_\mathrm{a}$, where $n_\mathrm{a}$ is the number density of atmospheric ions (H, He, electrons).\footnote{
For completeness, when $n_Z \sim n_\mathrm{a}$, the collision frequencies between metal ions will become important.
Eq. \eqref{eqn:ion_ion} shows that collisions with electrons will still dominate by orders of magnitude and ion-ion collisions can be safely ignored in general.}
If this condition is no longer true, such as when the instantaneous accretion rate is very high, then the opacity of these metals will affect atmospheric structures and their various scale heights.
In addition, there will be an additional term in the sinking timescale calculation due to the gradient of the metal concentration \citep{Burgers1969, Paquette1986, Pelletier1986}.
In this circumstance, new simulations must be done to assess the sinking timescale. 
Since the measured metal pollution rates may be underestimated by orders of magnitude, these effects may be important in the polluted patches of a white dwarf's atmosphere. 

Seventh, the surface covering fraction in Fig. \ref{fig:horizontal_spreading} assumed that the accretion is in a steady-state.
However, this has been suggested to be a poor approximation for many white dwarfs (e.g., see Appendix D in \citealt{Brouwers2023}).
Episodic accretion would produce a highly concentrated initial region that spreads across the white dwarf's surface over time. Ergo, it may be possible to detect variability even on white dwarfs with significant convection, although the probability of such an observation would be reduced in comparison to a white dwarf with a small steady-state surface fraction.
While it is possible to solve the full time-dependent diffusion-advection equation (Eq. \ref{eqn:full_eqn}), given the significant uncertainties in both horizontal and vertical timescales discussed above, we defer this investigation to future studies.

\section{Conclusions}\label{sec:conc}

In this paper, we have produced models of magnetic accretion of metals onto polluted white dwarfs and the subsequent evolution of these metals in the atmosphere, as well as the observational impacts of these processes. 
These models are summarized in Figs. \ref{fig:systemdiagram} and \ref{fig:atmosphere_diagram} and briefly below.
First, the debris disk around white dwarfs is assumed to be the result of a tidal disruption event, with some inclination relative to the white dwarf's spin pole. 
The structure of this disk, including the dust sublimation radius and the radii where the magnetic field will accrete material, are described (Sec. \ref{sec:accdisk}). 
Once the material is coupled to the magnetic field, it will be accreted onto the white dwarf's surface in a narrow beam. The geometry and area of this beam is discussed (Sec. \ref{sec:spatialdist}). 
This model is amenable to a wide range of disk geometries and is used to study the effects of this geometry on the incoming beam. 
Then, we model the atmospheric dynamics of the polluting beam, including the stopping distance of the beam through the atmosphere, the collision and gyro-frequencies of the metals, and the relevant vertical sinking and surface horizontal spreading timescales (Sec. \ref{sec:diff}).
These calculations allow us to find the area of the pollution patch on white dwarf atmospheres across parameter space and to determine sufficient conditions for globalized and homogeneous pollution (Fig. \ref{fig:horizontal_spreading}, Sec. \ref{sec:covering_with_beam}).
Finally, the effects of a polluted patch on observational signatures such as the variability and the inferred accretion rate are calculated in Sec. \ref{sec:obs}. 

Our main results can be summarized as follows:
\begin{enumerate}
    \item Various radii describing the accretion are calculated in Fig. \ref{fig:radiicomp}. First a body is tidally disrupted at the Roche radius. A dust disk is created, although this dust disk can never be magnetically accreted. As the dust drifts inward it will be sublimated, and then magnetically accreted along field lines. This magnetic accretion will occur even when the magnetic field strength is as low as \qty{1}{kG}. 
    \item The debris disk's orientation relative to the magnetic pole can produce orders of magnitude difference in the polluting beam size, with circumpolar disk producing the maximum beam size (Fig. \ref{fig:coverfrac}).
    \item Although the incoming velocity of the polluting beam is high, it cannot punch through a significant fraction of the atmosphere. Once a polluting ion is in the white dwarf's atmosphere, its motion is dominated by collisions with electrons rather than magnetic gyration.
    \item The metal surface covering fraction on white dwarfs (Fig. \ref{fig:horizontal_spreading}) span many orders of magnitude and are strongly dependent on temperature and the presence of convection. These results are calculated through the ratio $\tau_\mathrm{sink}/\tau_\mathrm{spread}$ and can be adapted for future models of horizontal and vertical diffusion timescales.
    \item The model we present here demonstrates that the variability of WD 2138-332 could be explained as a surface metal spot approximately 10\% of the white dwarf's surface area.
    \item These models demonstrate that variability in the accretion signatures can only occur if the spin and magnetic pole axes are misaligned. Patchy pollution is therefore difficult to detect. 
    \item Polluted white dwarfs with unresolved patches may have true accretion rates up to a few orders of magnitude larger than measured due to both the metal spot and the effects of multiple polluting elements.
    \item Full three-dimensional models of the atmosphere of a magnetic white dwarf are necessary to better understand pollution processes and the impacts of magnetically-channeled accretion.
\end{enumerate}

\section*{Acknowledgments}

We thank the two anonymous reviewers for their helpful and detailed comments.
We thank Pier-Emmanuel Tremblay for his insightful comments on the effects of magnetic fields on convection in white dwarfs and JR Fuentes for his assistance with the effects of magnetic fields on the convective turnover time.
We thank Fred Adams, Steve Majeski, Nuria Calvet, Lee Hartmann, Evan Bauer, Mitchell Begelman, Chris Matzner, and Alexander Mule for helpful conversations.
D.P. acknowledges support from the McCray Postdoctoral Fellowship at the University of Colorado Boulder.
A.G.T. acknowledges support from the Fannie and John Hertz Foundation and the University of Michigan's Rackham Merit Fellowship Program.

\appendix
\section{Surface Accretion Point}\label{appendix:accpt}

This section derives the location of the accretion point on the surface of a white dwarf for a highly eccentric debris disk with accretion from a single point in the disk. 
Consider an offset angle $\beta$ between the magnetic field pole and the spin axis of the white dwarf and an angle $\phi$ between the magnetic field pole and the location in the disk where the material is accreted from. We can consider this problem in the  corotating frame of the white dwarf and choose $\phi$ such that $\phi=0$ when the spin pole, magnetic field pole, and accretion point are coplanar. 

Consider the Cartesian coordinate system where the spin pole is given by $\hat{\mathbf{s}}=\hat{\mathbf{z}}$. In this frame, the magnetic pole vector is
\begin{equation}
    \hat{\mathbf{B}}=\sin\beta \,\hat{\mathbf{x}}+\cos\beta\,\hat{\mathbf{z}}\,
\end{equation} 
and the direction to the accretion point is given by 
\begin{equation}
    \hat{\mathbf{p}}=\sin I\cos\phi\,\hat{\mathbf{x}}+\sin I\sin\phi\,\hat{\mathbf{y}}+\cos I\,\hat{\mathbf{z}}\,.
\end{equation}
In this equation, $I$ is the inclination between the disk and the white dwarf's spin axis. 

The angle $\theta$ between the accretion point and the magnetic field pole is then given by
\begin{equation}
    \cos\theta=\hat{\mathbf{B}}\cdot\hat{\mathbf{p}}=\sin I\sin\beta\cos\phi+\cos\beta\cos I\,.
\end{equation}
Eq. \eqref{eq:theta0cond} is then used to solve for $\theta_0$, the location where the accretion flow strikes the white dwarf, as a function of $\beta$, $I$, and $\phi$.

The direction where the accreted material strikes the white dwarf's surface, characterized by the unit vector $\hat{\mathbf{u}}$, must (i) be in the plane spanned by $\hat{\mathbf{B}}$ and $\hat{\mathbf{p}}$ and (ii) satisfy $\hat{\mathbf{u}}\cdot\hat{\mathbf{B}}=\cos\theta_0$. These conditions will be satisfied by 
\begin{equation}
    \hat{\mathbf{u}}=\cos\theta_0\,\hat{\mathbf{B}}+\frac{\sin\theta_0}{\sin\theta}\left(\hat{\mathbf{p}}-\cos\theta\,\hat{\mathbf{B}}\right)\,.
\end{equation}
This vector is a linear combination of $\hat{\mathbf{B}}$ and $\hat{\mathbf{p}}$ and so is in the plane spanned by these vectors, and is constructed so that the second condition is satisfied. 

In the corotating frame, the position angles $\varphi$, $\vartheta$ of the accretion region are therefore given by $\vartheta=\tan^{-1}(\hat{\mathbf{u}}_z/\sqrt{\hat{\mathbf{u}}_x^2+\hat{\mathbf{u}}_y^2})$ and $\varphi=\tan^{-1}(\hat{\mathbf{u}}_y/\hat{\mathbf{u}}_x)$, with appropriate corrections for the domain of the azimuthal angle $\varphi$. Here, $\varphi$ is the azimuthal angle and $\vartheta$ is the polar angle. 

For a rotating white dwarf, the phase $\phi$ of the accretion point will rotate from 0 to $2\pi$, leading to variable accretion locations on the surface of the white dwarf. This can be accounted for by solving the above equations to find $\varphi$ and $\vartheta$ for $\phi$ values across the entire domain. 

\section{Stopping Length}\label{appendix:stoppinglength}

This section derives the stopping length of the accretion flow of metals into the photosphere. We show that the accreting material is stopped in a fraction of the photosphere and does not penetrate into the white dwarf's surface. 

Since the incoming velocity $v_{\rm in}$ is very high (Eq. \ref{eqn:vin}), the thermal energy of the incoming particle is much less than the kinetic energy for any relevant ionization temperature ($\sim 10^4 - 10^5$ K).
Thus, the energy of the incoming particle is
\begin{align}
    E_\mathrm{in} \approx &~\frac{1}{2}m_Z v_\mathrm{in}^2 \nonumber\\
    \simeq &~ \qty{4.8e-6}{erg} \left(\frac{m_Z}{m_\mathrm{Ca}}\right) \left(\frac{v_\mathrm{in}}{\qty{4.8e8}{cm~~s^{-1}}}\right)^2.
\end{align}
Here, $m_Z$ is the mass of the incoming particle.

As the incoming particle interacts with particles in a white dwarf atmosphere, its velocity is decelerated as \citep[see NRL Plasma Formulary, ][]{Huba2013}:
\begin{equation}
    \frac{\de v}{\de t} = -\nu_\mathrm{slow} v,
\end{equation}
where $\nu_\mathrm{slow}$ is due to two sources --- atmospheric electrons and atmospheric ions.
Due to interactions with electrons (in the very fast limit), $\nu_\mathrm{slow}$ is
\begin{align}\label{eqn:nu_e}
    \nu_\mathrm{slow, e} \approx&~ \qty{2.0e5}{s^{-1}} \left(\frac{m_Z}{m_\mathrm{Ca}}\right)^{1/2} \left(\frac{Z}{2}\right)^2 \times\nonumber\\ &~\left(\frac{E_\mathrm{in}}{\qty{4.8e-6}{erg}}\right)^{-3/2} \left(\frac{n_e}{\qty{5e16}{cm^{-3}}}\right) \times\nonumber\\
    &~\left(\frac{\ln \Lambda_{i,e}}{10}\right)\,,
\end{align}
where is the charge state of the incoming ion (assumed to be $^{40}$Ca), $n_e$ is the electron number density in the photosphere, and $E_{\rm in}$ is the kinetic energy of the incoming particles.
The Coulomb logarithm for ion-electron collisions $\ln \Lambda_{i,e}$ has typical values on the order of 10.
Likewise, $\nu_\mathrm{slow}$ due to interactions with atmospheric hydrogen ions is
\begin{align}
    \nu_\mathrm{slow, H} \approx&~ \qty{1.1e2}{s^{-1}} \left(\frac{m_Z}{m_\mathrm{Ca}}\right)^{1/2} \left(\frac{Z}{2}\right)^2 \left(\frac{Z_H}{1}\right)^2 \nonumber\times\\
    &~\left(\frac{E_\mathrm{in}}{\qty{4.8e-6}{erg}}\right)^{-3/2} \left(\frac{n_e}{\qty{5e16}{cm^{-3}}}\right) \times\nonumber\\
    &~\left(\frac{\ln \Lambda_{i,i}}{10}\right),
\end{align}
where $Z_H$ is the ionization state of the atmospheric ion and $\ln \Lambda_{i,i}$ is the ion-ion Coulomb logarithm.
This expression also assumes that the incoming ion has a much greater mass than the atmospheric ion, which holds for most polluting elements of interest in the literature (e.g., Ca, Si, Fe) onto hydrogen and helium atmosphere white dwarfs.

The stopping length is then given by
\begin{equation}
    L_\mathrm{stop} \sim \frac{v_\mathrm{in}}{\nu_\mathrm{slow}}.
\end{equation}
As previously calculated, $\nu_\mathrm{slow, e} \gg \nu_\mathrm{slow, H}$, giving us $\nu_\mathrm{slow} = \nu_\mathrm{slow, e}$.

\section{The Diffusion Equation}\label{appendix:diffusion}

In this section, we derive the surface abundance of polluting metals across the white dwarf's surface for horizontal diffusion from a narrow accretion region. This result sets the size of the pollution region in terms of the horizontal diffusion and vertical settling timescales. 

\subsection{Green's Function}

The full time-dependent drift-diffusion equation describing the horizontal spreading and vertical sinking concentration of polluting particles is \citep[cf.][]{Cunningham2021}
\begin{equation}\label{eqn:full_eqn}
    \frac{\partial u}{\partial t} = D_\mathrm{surf}\nabla^2u - v_\mathrm{drift} \frac{\partial u}{\partial z},
\end{equation}
where $u = u(t,x,y,z)$ is the concentration (per unit area) of the trace metal as a function of time and position, $v_{\rm drift}$ is the average vertical sinking velocity, and $D_\mathrm{surf}$ is the horizontal spreading diffusion coefficient.

In the case of constant accretion from the disk onto the white dwarf, we seek to model the surface spreading over coordinate $(x,y)$ and vertical sinking over the depth $z$.
Then, the time dependence drops and Eq. \eqref{eqn:full_eqn} equation becomes
\begin{equation}
    D_\mathrm{surf}\nabla^2 u = v_\mathrm{drift}\partial_z u\,,
\end{equation}
where we have used the notation $\partial_z u = \partial u/\partial z$.
The Laplacian operator is acting on $u$ in $(x,y)$ so that
\begin{equation}
    D_\mathrm{surf}\left(\partial_x^2 u + \partial_y^2u\right) -v_\mathrm{drift}\partial_z u = 0\,.
\end{equation}
This equation is the two-dimensional heat equation in $(x,y)$ with $z$ instead of the usual time coordinate.

First, we find the solution to this equation when accretion constantly arrives at a single spot $(x,y,z)=(0,0,0)$, so that the equation becomes
\begin{equation}
    D_\mathrm{surf}\left(\partial_x^2 u + \partial_y^2u\right) -v_\mathrm{drift}\partial_z u = \delta(x)\delta(y)\delta(z)\,,
\end{equation}
where $\delta$ is the Dirac delta function.
Solving this equation can be done in various ways and the solution is well-known. It takes the form
\begin{equation}
    u_\delta(\mathbf{x}, z) = \frac{v_\mathrm{drift}}{4\pi D_\mathrm{surf}} \cdot z^{-1} \exp\left(-\frac{|\mathbf{x}|^2}{4z}\frac{v_\mathrm{drift}}{D_\mathrm{surf}}\right)
\end{equation}
where we have defined
\begin{equation}
    \mathbf{x} = (x,y) ~\mathrm{and}~ |\mathbf{x}|^2 = x^2 + y^2.
\end{equation}
Note that the exponential term is in the form of a Gaussian with variance
\begin{equation}
    \sigma^2 \equiv 2 D_\mathrm{surf}\frac{z}{v_\mathrm{drift}}.
\end{equation}
Finally, this solution is equivalent to finding the Green's function $G(\mathbf{x}, z)$ such that $G(\mathbf{x},\mathbf{x}', z) \equiv u_\delta(\mathbf{x}-\mathbf{x}', z)$.

\subsection{Surface Pollution}
The previously found solution can be used to derive the conditions under which the pollution zone will cover the entire atmosphere, so that the metal concentration at the far side of the star ($x^2+y^2=R_\star^2$) and at the bottom of the convection zone/photosphere ($z=z_{\rm cvz}$) --- given by $u(x^2+y^2=R_*^2, z_{\rm cvz})$ --- has the same order of magnitude as $u(0, 0, 0)$, the concentration under the pollution beam. 
With this condition, we can define the horizontal spreading timescale as
\begin{equation}
    \tau_\mathrm{spread} \equiv \frac{R_*^2}{4 D_\mathrm{surf}}.
\end{equation}
The remaining quantity in the exponential term is $\tau_\mathrm{vert} = z/v_\mathrm{drift}$, which acts as the timescale for vertical advection from the top of atmosphere at $z=0$ to some depth $z$.
Here, we are interested in the lifetime of an ion in the observable reservoir, $0 \leq z \leq z_\mathrm{cvz}$, where metal spreading can still affect the surface pollution spot size.
Metals are transported to the bottom of the convection zone on a timescale $\tau_\mathrm{vert}(z_\mathrm{cvz})=z_\mathrm{cvz}/v_\mathrm{drift}$, while simultaneously being lost through $z_\mathrm{cvz}$ on the sinking timescale $\tau_\mathrm{sink}$ (Eq. \ref{eqn:sinking_timescale}).
Since we have a steady-state model, the downward flux to $z_\mathrm{cvz}$ must match the rate at which materials are removed from this layer:
\begin{equation}
    \tau_\mathrm{vert}(z_\mathrm{cvz}) = \tau_\mathrm{sink}.
\end{equation}
This boundary condition is necessary to ensure that materials do not accumulate at the bottom of the convection zone, which would violate the steady-state assumption.

With these timescales defined, the solution takes the form
\begin{equation}
    u(R_\star, \tau_\mathrm{sink})\propto(D_\mathrm{surf}\tau_\mathrm{sink})^{-1}\exp\left(-\frac{\tau_\mathrm{spread}}{\tau_\mathrm{sink}}\right)\,
\end{equation}
which describes the pollution gradient over the sinking timescale $t_\mathrm{sink}$.

Alternatively, by recognizing that the exponential term in the solution is a two-dimensional Gaussian with variance
\begin{equation}\label{eqn:variance_spot}
    \sigma^2 \equiv 2 D_\mathrm{surf} \tau_\mathrm{sink}
\end{equation}
and again considering $x^2+y^2=R_\star^2$, the ratio between the \qty{1}{\sigma} polluted area and the total white dwarf surface area can be estimated as
\begin{equation}\label{eq:fcovspread}
    f_\mathrm{cov} \simeq \frac{\pi\sigma^2}{4\pi R_*^2} = \frac{1}{8} \frac{\tau_\mathrm{sink}}{\tau_\mathrm{spread}}.
\end{equation}

\subsection{Pollution with an Incoming Beam}

Now, we seek to find the solution when the incoming beam arriving at $z=0$ is a Gaussian described by a variance $\sigma_\mathrm{beam}$ and takes the form
\begin{equation}
    u_0(\mathbf{x}';z=0) =\left(\frac{C_0}{2\pi\sigma_\mathrm{beam}^2}\right) \exp\left(-\frac{|\mathbf{x'}|^2}{2 \sigma_\mathrm{beam}^2}\right)
\end{equation}
where the normalization constant $C_0$ is defined such that $C_0$ is the total amount of material constantly being delivered by the beam,
\begin{align}
    C_0 = \int \de \mathbf{x}'~ u_0(\mathbf{x}').
\end{align}

The solution for this particular initial condition can be readily found by convolving the source term with the Green's function, so that 
\begin{align}
    u_\mathrm{b}(\mathbf{x},z) &= \int \!\!\de \mathbf{x}'\,G(\mathbf{x},\mathbf{x}',z)\, u_0(\mathbf{x}')\nonumber\\
    &= \left(\frac{C_0}{2\pi\sigma_\mathrm{eff}^2}\right)\exp\left(-\frac{|\mathbf{x}|^2}{2\sigma_\mathrm{eff}^2}\right)
\end{align}
where the new variance is defined as
\begin{align}
    \sigma_\mathrm{eff}^2(z) &\equiv \sigma^2(z)  + \sigma_\mathrm{beam}^2 \nonumber\\
    &= 2 D_\mathrm{surf}\frac{z}{v_\mathrm{drift}} + \sigma_\mathrm{beam}^2.
\end{align}
Since both the Green's function and our initial condition are Gaussians, the convolution of the two (the solution) is also expected to be a Gaussian. 
Finally, we now find $f_\mathrm{cov}$, the ratio between the $1\sigma$ polluted area and the total white dwarf surface area, with an incoming beam having a width $\sigma_\mathrm{beam}$ as
\begin{equation}\label{eq:fcovtot}
    f_\mathrm{cov} \simeq \frac{1}{8}\frac{\tau_\mathrm{sink}}{\tau_\mathrm{spread}} + \frac{\sigma_\mathrm{beam}^2}{4R_*^2}.
\end{equation}

\bibliography{main}{}
\bibliographystyle{aasjournal}

\end{document}